\documentclass[aps,prx,twocolumn,superscriptaddress,longbibliography,showpacs,floatfix,10pt]{revtex4-2}

\usepackage{amsmath}
\usepackage{graphicx}
\usepackage{color}
\usepackage{epstopdf}
\usepackage{natbib}
\usepackage{xfrac}
\usepackage{wasysym}
\usepackage{multirow}
\usepackage[normalem]{ulem}
\usepackage{amsfonts}

\usepackage{orcidlink}

\usepackage[margin=10pt,format=hang,caption=false,singlelinecheck=false]{subfig}

\usepackage{adjustbox} 

\usepackage{xcolor,colortbl}
\definecolor{Gray}{gray}{0.85}
\definecolor{lblue}{rgb}{0.8,0.8,1}  %{blue}{0.5}
\definecolor{blue}{rgb}{0.6,0.6,0.9}  %{blue}{0.5}

\newcommand{\eref}[1]{Eq.~(\ref{#1})}
\newcommand{\fref}[1]{Fig.~\ref{#1}}

\newcommand{\sref}[1]{Section~\ref{#1}}
\newcommand{\aref}[1]{Appendix~\ref{#1}}
\newcommand{\tref}[1]{Tab.~\ref{#1}}
\newcommand{\vek}[1]{ \hbox{\textbf #1}}
\newcommand{\svek}{\mathbf}

\newcommand{\etal}{{\it et al.}}

\renewcommand{\Im}{\hbox{Im}}
\renewcommand{\Re}{\hbox{Re}}

\newcommand{\II}{\mathcal{I}}
\newcommand{\LL}{\mathcal{L}}
\newcommand{\KK}{\mathcal{K}}

\DeclareMathOperator{\sign}{sign}

\def\app#1#2{%
  \mathrel{%
    \setbox0=\hbox{$#1\sim$}%
    \setbox2=\hbox{%
      \rlap{\hbox{$#1\propto$}}%
      \lower1.1\ht0\box0%
    }%
    \raise0.25\ht2\box2%
  }%
}

\usepackage{float}

\begin{document}
\title{Prototypical many-body 
signatures in transport properties of semiconductors
}

\author{Matthias Pickem\,\orcidlink{0000-0002-0976-845X}}
\affiliation{Institute of Solid State Physics, TU Wien,  A-1040 Vienna, Austria}
\author{Emanuele Maggio\,\orcidlink{0000-0002-5528-5117}}
\affiliation{Institute of Solid State Physics, TU Wien,  A-1040 Vienna, Austria}
\author{Jan M.~Tomczak\,\orcidlink{0000-0003-1581-8799}}
\affiliation{Institute of Solid State Physics, TU Wien,  A-1040 Vienna, Austria}

\date{\today}

\begin{abstract}
We devise a methodology for charge, heat, and entropy transport driven by carriers with finite lifetimes.
Combining numerical simulations with analytical expressions for low temperatures, 
we establish a comprehensive and thermodynamically consistent phenomenology for transport properties in semiconductors. 
We demonstrate that the scattering rate (inverse lifetime) is a relevant energy scale:
It causes the emergence of several characteristic features in each transport observable. 
The theory is capable to reproduce---with only a minimal input electronic structure---the full temperature profiles 
measured in correlated narrow-gap semiconductors. In particular, we account for the previously elusive low-$T$ saturation 
of the resistivity and the Hall coefficient, as well as the (linear) vanishing of the Seebeck and Nernst coefficient in systems, such as FeSb$_2$, FeAs$_2$, RuSb$_2$ and FeGa$_3$.
\end{abstract}

\maketitle

\section{Introduction}
Transport properties---such as resistivity ($\rho$), magneto-resistance (MR), thermal conductance ($\kappa$), and the coefficients of Hall, Seebeck and Nernst ($R_H$, $S$, $\nu$)---are among the most widely 
investigated quantities in materials science.  They provide essential information for characterizing new materials and for elucidating 
physical phenomena.
To extract microscopic information from 
measurements requires a fundamental understanding of how carriers transport charge, heat and entropy.
When simulating transport properties, an adequate inclusion of scattering 
processes is particularly crucial. These limit the lifetime of carriers, lead to a decay of 
currents, and can have various origins,
such as
 electron-electron or electron-phonon interactions as well as defects or impurities. 

In this paper, we establish the prototypical signatures of finite electronic lifetimes in transport properties of (narrow-gap) semiconductors\cite{NGCS,Ponce_2020}.
To this end, we develop a methodology based on Kubo linear-response theory
which captures (in)coherence effects beyond the reach of semi-classical Boltzmann approaches\cite{Madsen200667,Boltztrap2,Pizzi2014422}, while incurring a comparable numerical cost.
Our theory reveals that at low enough temperatures, the
scattering rate becomes a relevant energy scale, in the sense that---contrary to Boltzmann theory---it determines the characteristic temperature profile rather than only scaling its amplitude. 
As a consequence, a rich {\it intrinsic} temperature dependence emerges in transport observables that previously has often been attributed to {\it extrinsic} effects.

\begin{figure}[H]
    \includegraphics[width=0.435\textwidth,clip=true,trim= 0 5 0 5]{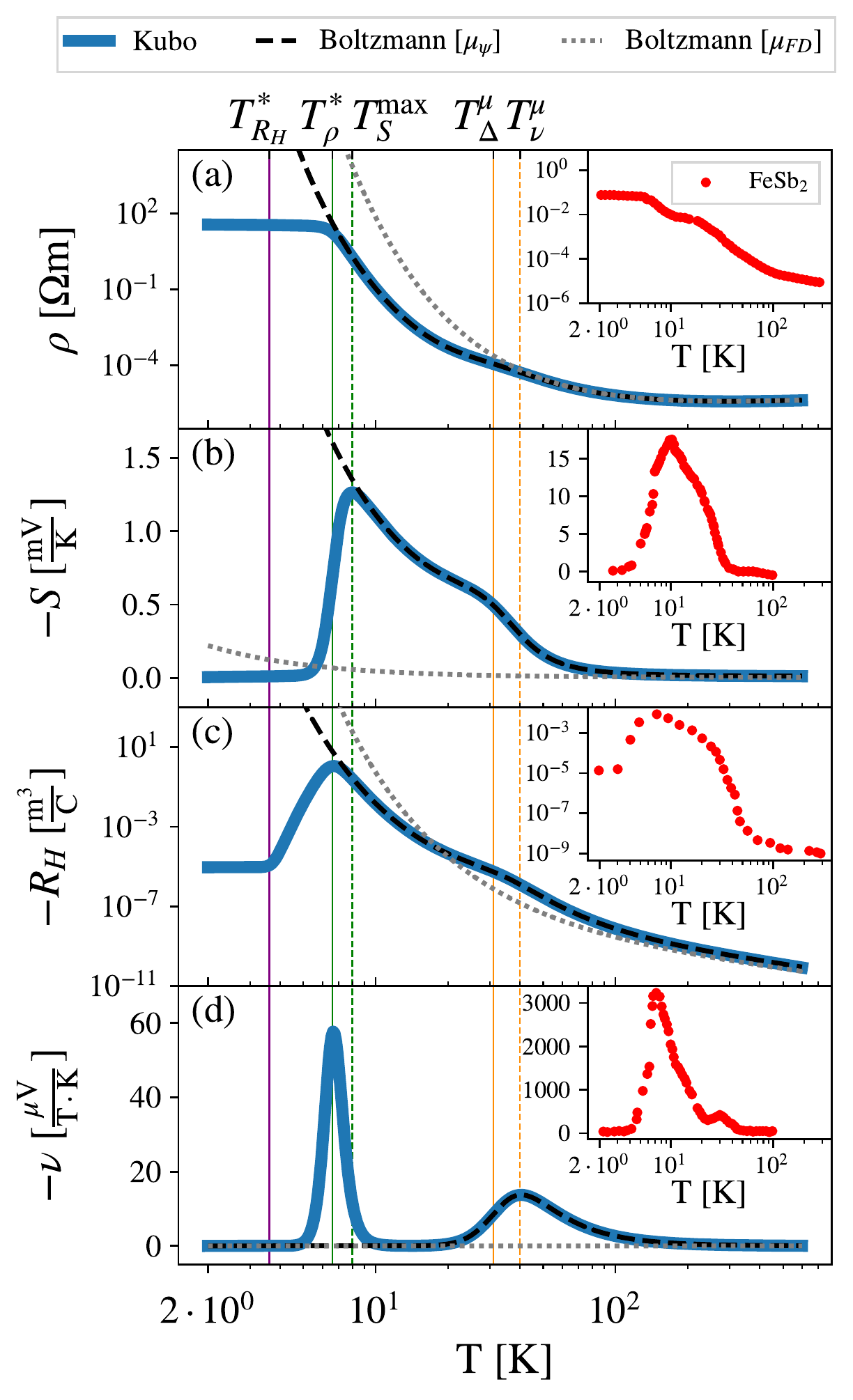} % l b r t
    \caption{\textbf{Prototypical transport in semiconductors.} Comparison of 
    a two-band model, $\epsilon^0_{\mathbf{k}n}$ $=$ $-\sum_{i=x,y,z}$  $2t_n \cos\left(k_i\right)$$+$$(-1)^n (6t_n$$+$$\Delta_0/2)$, with $t_1$$=$$250$meV, $t_2$$=$$-265$meV,
    bandgap \hbox{$\Delta_0=60$ meV}; effective mass \hbox{$Z^{-1}=2$}, scattering rate \hbox{$\Gamma(T) $$=$$ (5\cdot10^{-5} $$+$$ 10^{-7}/\hbox{K}^2 \cdot T^2)$ eV} with experiments for FeSb$_2$\cite{PhysRevB.88.245203} (insets):  a) resistivity, coefficients of b) Seebeck, c) Hall, and d) Nernst \emph{without} impurity in-gap states. Vertical lines mark characteristic temperatures (from left to right): Saturation onset of the Hall coefficient $T_{R_H}^*$ and the resistivity $T_\rho^*$, the maximal  Seebeck coefficient $T_S^\mathrm{max}$, onset of second activated regime $T^\mu_\Delta$, and the high-temperature Nernst peak $T^\mu_\nu$. Dashed black (grey) lines show Boltzmann results using a chemical potential, $\mu_{\psi}$ ($\mu_{FD}$), that accounts for lifetime and thermal (only thermal) broadening.
    }
    \label{fig:prototypical}
\end{figure}

We highlight this insight in \fref{fig:prototypical} for a simple two-band modeling of the colossal thermopower material FeSb$_2$\cite{0295-5075-80-1-17008,PhysRevB.72.045103,PhysRevB.88.245203,PhysRevResearch.2.023190}.
The large {\it magnitude} of its $S$ and $\nu$ 
originates
from the phonon-drag effect\cite{jmt_fesb2,MRC:8871060,Takahashi2016,FeSb2_Marco,doi:10.7566/JPSJ.88.074601,PhysRevB.103.L041202}%
\footnote{The phonon-enhancement of the electron diffusion is expected to be smooth in temperature, i.e., it does not introduce characteristic features. Further, a suppression of the phonon-drag in polycrystalline samples yields Seebeck coefficients\cite{bentien:205105,MRC:8871060,Sanchela2015205,Takahashi2016} comparable to our modelling.}.
Here, we focus on the presence of
{\it characteristic temperatures} that mark features
across various transport quantities\cite{PhysRevB.88.245203}: 
For instance, at low temperatures, inflection points in the resistivity $\rho$ and the Seebeck coefficient $S$ correlate with maxima in the Hall and Nernst coefficient, $R_H$, $\nu$. %
This intriguing---but by no means uncommon\cite{NGCS,sun_dalton,PhysRevB.90.195206}---temperature profile, has previously been advocated to derive from {\it extrinsic} in-gap states\cite{PhysRevB.88.165205,FeSb2_Marco,doi:10.7566/JPSJ.88.074601,PhysRevB.103.L041202,doi:10.1063/5.0048165,Du2021}.

Here, instead, we reproduce all qualitative temperature features in FeSb$_2$ exclusively by endowing the {\it intrinsic} valence and conduction carriers with a {\it finite scattering rate}, $\Gamma(T)=\Gamma_0+\gamma T^2$. 
In this scenario, possible impurity states influence transport 
solely by limiting the lifetime of intrinsic carriers through scattering,
not by providing additional carriers. 
Our findings establish a
new phenomenology for transport properties in semiconductors:
Below a temperature $T_\rho^*$, $\rho$ saturates\cite{LRT_Tstar}
instead of growing exponentially (see the Boltzmann result (dashed line) in \fref{fig:prototypical}).
$R_H$ also saturates (below $T_{R_H}^*<T_{\rho}^*$), indicating that residual scattering leads to a finite density of states even at absolute zero. Finite lifetimes also mend the violation of the 3rd law of thermodynamics of Boltzmann's relaxation time approximation: Instead of diverging, the Seebeck coefficient $S$ vanishes linearly for $T\rightarrow 0$.
Combined to the powerfactor $S^2\sigma$ and the figure of merit $zT$, our findings have practical relevance for thermoelectric applications: In narrow-gap semiconductors, these quantities exhibit large values at intermediate to low temperatures when scattering processes are properly accounted for. Material surveys based on Boltzmann approaches for coherent band structure instead fail to even qualitatively describe  $S^2\sigma$ and $zT$.
Finally, in congruence with experiment, a sharp low-$T$ feature emerges in the Nernst coefficient that, again, has no analogue in the Boltzmann treatment.

Looking at the available experimental literature, we find
a number of narrow-gap semiconductors\cite{NGCS} that exhibit qualitatively similar temperature profiles as the ones displayed in \fref{fig:prototypical}. For instance, other marcasite compounds (FeAs$_2$, RuSb$_2$\cite{APEX.2.091102,sun_dalton,1742-6596-150-1-012049,cava2013fesb2}, 
CrSb$_2$\cite{PhysRevB.86.235136}), silicides 
(FeSi\cite{Wolfe1965449,PhysRevB.83.125209,jmt_fesi}, 
RuSi\cite{Hohl199839,BUSCHINGER199757,jmt_hvar}), Heusler systems (e.g., Fe$_2$VAl\cite{PhysRevLett.79.1909,0953-8984-12-8-318,0953-8984-28-28-285601,doi:10.1143/JPSJ.74.1378,Hinterleitner2019}), other intermetallic compounds\cite{Likhanov2020} (e.g.,
FeGa$_3$, RuGa$_3$\cite{JPSJ.78.013702,monika_fega3,PhysRevB.90.195206}), as well as Kondo insulators (e.g., 
Ce$_3$Bi$_4$Pt$_3$\cite{PhysRevB.42.6842,PhysRevB.55.7533,Katoh199822,PhysRevB.94.035127,PhysRevB.100.235133,LRT_Tstar,jmt_CBP_arxiv})---strongly 
suggesting that our scenario based on carriers with finite lifetimes is prototypical for a wide array of different systems.

The paper is organized as follows: After introducing the formal background in \sref{sec:methcontext}, we present the methodological advances of our approach in \sref{sec:methadv}. 
Then, in \sref{sec:results}, we set out to establish a new phenomenology of transport properties of (narrow-gap) semiconductors:
In \sref{sec:lowT} we provide analytical results for the low temperature asymptotic behavior of all considered transport functions. In \sref{sec:parameterscan} we survey how the size of the charge gap $\Delta$, the particle-hole asymmetry $\alpha$, and the scattering rate $\Gamma$ control the temperature profile of observables. Crucially, we elucidate how finite lifetimes conspire with $\Delta$ and $\alpha$ to cause the emergence of several characteristic energy scales that appear in all transport observables.
In \sref{sec:results2}, we discuss implications for materials simulations on the basis of a simple modelling of selected intermetallic narrow-gap semiconductors\cite{NGCS}.
Finally, we end with a summary in \sref{sec:summary}.

\section{Methodological Context}\label{sec:methcontext}
In this section, we lay out the setting in which we consider transport properties.
The purpose is mainly to introduce the considered transport quantities and some necessary notation. For more detailed derivations, the reader is referred to specialized literature \cite{mahan,zlatic,behnia}, as well as Refs.~\onlinecite{Wenhu_thesis,Pickem}.
The conceptual advances beyond this setting will be presented in \sref{sec:methadv}, with more details in \aref{sec:app_kernels}.
The experienced reader
may jump to the 
analytical results for low temperatures, \sref{sec:lowT}, or directly to the numerical
results in Section \ref{sec:results}.

\subsection{Linear Response}
In linear response, transport quantities are based on correlation functions that specify measurable observables of a system in the presence of specific external perturbations (electric field, magnetic field, temperature gradient, etc.).
In our case these processes are described, on the imaginary time ($\tau$) axis by
\begin{equation}
    \chi_{\mathbf{j}^a\mathbf{j}^b}^{\alpha\beta}(\mathbf{q},\tau) = \frac{1}{V} \left\langle \mathcal{T}_\tau {j}^a_\alpha(\mathbf{q},\tau) {j}^b_\beta(-\mathbf{q},0)\right\rangle
    \label{chi}
\end{equation}
with the time-ordering operator $\mathcal{T}_\tau$, the charge ($a$, $b=1$) and heat ($a$, $b=2$) current operator $\mathbf{j}^{a}_{\alpha}$ in the Cartesian direction $\alpha$, $\beta\in\{x,y,z\}$, and $V$ indicating the unit cell volume.
From them, the usual (retarded) Onsager coefficients $\mathcal{L}$
for dipolar transitions ($\mathbf{q}=0$)
are obtained by first Fourier transforming \eref{chi} into bosonic Matsubara frequencies
\begin{equation}
    \chi_{\mathbf{j}^a\mathbf{j}^b}^{\alpha\beta}(\mathbf{q},i\omega_n) = \int_0^\beta d\tau e^{i\omega_n \tau} \chi_{\mathbf{j}^a\mathbf{j}^b}^{\alpha\beta}(\mathbf{q},\tau),
\end{equation}
analytical continuation to real frequencies $i\omega_n \rightarrow \omega + i\delta$ and then taking the dynamic limit 
\begin{equation}
    \mathcal{L}_{ab}^{\alpha\beta} = \lim_{\omega\rightarrow 0^+} \frac{1}{\omega} \Im \left[ \chi_{\mathbf{j}^a\mathbf{j}^b}^{\alpha\beta}(\mathbf{q}=0, \omega)\right].\label{onsager}
\end{equation}
In the presence of an external magnetic field $B$ in direction $\gamma\in\{x,y,z\}$, 
one needs to instead evaluate the expectation value
\begin{equation}
\chi_{\mathbf{j}^a\mathbf{j}^b}^{B,\alpha\beta\gamma}(\mathbf{q},\tau) = \frac{1}{V} \left\langle T_\tau {j}^a_\alpha(\mathbf{q},\tau) {j}^b_\beta(\mathbf{0},0)\right\rangle_{B_\gamma} \label{chiB}
\end{equation}
for the Hamiltonian that includes the field\cite{PhysRevB.45.13945,10.1143/PTP.80.623,Wenhu_thesis}, and the resulting Onsager coefficients will be denoted
\begin{equation}
    \mathcal{L}_{ab}^{B,\alpha\beta\gamma} = \lim_{\omega\rightarrow 0^+} \frac{1}{\omega} \Im \left[ \chi_{\mathbf{j}^a\mathbf{j}^b}^{B,\alpha\beta\gamma}(\mathbf{q}=0, \omega)\right].\label{onsagerB}
\end{equation}
From the above, the observable transport tensors can be derived.
Namely, the conductivity $\sigma$, the resistivity $\rho$, the thermopower (Seebeck coefficient) $S$, the electronic thermal conductivity $\kappa$, the Hall conductivity $\sigma^B$, the Hall coefficient $R_H$, and the Nernst coefficient $\nu$:

\begin{align}
\sigma_{\alpha\beta}&= \mathcal{L}_{11}^{\alpha\beta} \label{eq:ObservCond}\\
\rho_{\alpha\beta}&= \left(\mathcal{L}_{11}^{-1}\right)^{\alpha\beta}\label{eq:ObservRes}\\
S_{\alpha\beta}&=-\frac{1}{T} \left(\LL_{11}^{-1}\right)^{\alpha i}\LL_{12}^{i\beta} \label{eq:ObservSeeb}\\
\kappa_{\alpha\beta}&= \frac{1}{T}\left[ \mathcal{L}_{22}^{\alpha\beta} -   \LL_{12}^{\alpha  i} \left(\LL_{11}^{-1}\right)^{ij} \LL_{12}^{j\beta}\right] \label{eq:ObservThermal}\\
\sigma^B_{\alpha\beta\gamma}&= \mathcal{L}_{11}^{B,\alpha\beta\gamma} \label{eq:ObservHallCond}\\
R_{H,\alpha\beta\gamma}&= \left(\LL_{11}^{-1}\right)^{\alpha i}
\LL_{11}^{B,ij\gamma} \left(\LL_{11}^{-1}\right)^{j \beta} \label{eq:ObservHallCoeff}\\
\nu_{\alpha\beta\gamma}&=-\frac{1}{T} \left(\LL_{11}^{-1}\right)^{\alpha i} \left[ \LL_{11}^{B,ij\gamma} \LL_{12}^{jk} - \LL_{12}^{B,ij\gamma} \LL_{11}^{jk}\right] \left(\LL_{11}^{-1}\right)^{k\beta}
\label{eq:ObservNernst}
\end{align}

For model/materials whose unit cell's axes are orthogonal, as we are considering here, the Einstein summations over Cartesian directions simplify to a single expression, e.g., for an electric current in $x$-direction and a magnetic field in $z$-direction, the resulting Hall coefficient in $y$-direction is $R_{H,xyz} = \left(\LL_{11}^{-1}\right)^{xx}
\LL_{11}^{B,xyz} \left(\LL_{11}^{-1}\right)^{yy}$.
Later, we will also address the (empirical) Wiedemann-Franz law
\begin{equation}
L=\frac{\kappa}{\sigma T} 
\label{eq:Lorenz}
\end{equation}
as well as the
thermoelectric
power factor
\begin{equation}
    PF = S^2 \sigma
\label{eq:PF}
\end{equation}
and figure of merit
\begin{equation}
   zT = \frac{S^2 \sigma T}{\kappa}.
\label{eq:FOM}
\end{equation}

\subsection{Integral equations}
\subsubsection{One vs.\ multi-particle correlations}
The transport observables, even when featuring one-particle currents $\mathbf{j}^a_\alpha$ \footnote{
The heat-current $\mathbf{j}^2$ is only of one-particle nature when, as we assume here, interactions are local\cite{PhysRevB.67.115131}.
} 
in \eref{chi} and \eref{chiB}, probe multi-particle excitations. Diagrammatically, these can be described as
the sum of all possible two-particle processes, with the leading term corresponding to the independent propagation of a particle-hole pair (bubble diagram).
Magneto-transport quantities on the other hand stemming from \eref{chiB} can be shown\cite{10.1143/PTP.80.623,PhysRevB.45.13945,Wenhu_thesis,Pickem} to involve all possible three-particle processes.
Diagrams that (unlike the bubble) contain interconnected propagation lines, are commonly referred to as vertex-corrections\cite{kontani,PhysRevB.67.014408,PhysRevB.98.165130,PhysRevLett.123.036601}. 
These can lead to collective phenomena such as excitons, $\pi$-tons\cite{PhysRevLett.124.047401,PhysRevB.103.104415,PhysRevB.104.115153} and other polaritons.
In this work, following the spirit of the dynamical mean-field theory\cite{bible}, we are neglecting vertex-corrections. In this approximation, which amounts to the infinite dimensional limit,
vertex corrections vanish for all considered observables \cite{PhysRevLett.64.1990,bible,PhysRevB.67.115131,PhysRevB.100.115102}
\footnote{at least in the absence of multi-band effects\cite{me_phd}.
For the vanishing of vertex corrections in infinite dimensions for massless fermions, see Ref.~\onlinecite{PhysRevLett.126.206601}.}.

Assuming that the one-particle Green's function $G_{\mathbf{k}}(\omega)$ is diagonal in the chosen band or orbital basis, the Onsager coefficients, Eqs.~(\ref{onsager}-\ref{onsagerB}), can be written as
\begin{align}
    \mathcal{L}_{ab}^{\alpha\beta} &= \frac{\pi\hbar e^{\left(4-a-b\right)} }{V} \sum_{\substack{n,m\\ \mathbf{k},\sigma}} \mathcal{K}_{ab}(\mathbf{k},n,m) M^{\alpha\beta}(\mathbf{k},n,m)\label{eqL}\\
    \mathcal{L}_{ab}^{B,\alpha\beta\gamma} &= \frac{4\pi^2 \hbar e^{\left(5-a-b\right)}}{3V} \sum_{\substack{n,m\\ \mathbf{k},\sigma}} \mathcal{K}_{ab}^B(\mathbf{k},n,m) M^{B,\alpha\beta\gamma}(\mathbf{k},n,m)\label{eqLB}
\end{align}
with the electron charge $e$, and the sums running over band-indices $n$, $m$, Brillouin zone momentum $\mathbf{k}$ and spin $\sigma$.
Here, the $M^{(B)}$ collect the
 dipolar transition matrix elements that depend on the Cartesian directions $\alpha$, $\beta$ (and $\gamma$) and are given in
 the next paragraph. 
The kernel functions $\mathcal{K}^{(B)}$, instead, 
contain the two(three)-particle expectation value of the 
fermionic operators that make up the currents $\mathbf{j}^a$.
Neglecting vertex corrections (see above), they can be expressed as
\begin{align}
    \mathcal{K}_{ab}(\mathbf{k},n,m) &= \int_{-\infty}^{\infty}d\omega\; \omega^{(a+b-2)} \left( - \frac{\partial f}{\partial\omega} \right) A_{\mathbf{k}n}(\omega) A_{\mathbf{k}m}(\omega)
    \label{K1}\\
    \mathcal{K}_{ab}^{B}(\mathbf{k},n,m) &= \int_{-\infty}^{\infty}d\omega\;  \omega^{(a+b-2)} \left( - \frac{\partial f}{\partial\omega} \right) A^2_{\mathbf{k}n}(\omega)A_{\mathbf{k}m}(\omega) 
    \label{K2}
\end{align}
where $A_{\mathbf{k}n}(\omega)=-1/\pi\Im G_{\mathbf{k}n}(\omega)$ is the spectral function associated with the retarded one-particle Green's function. Energies $\omega$ are measured with respect to the Fermi level $\mu$.
Thus, within our approximations, 
many-body (scattering) effects enter the transport properties only through the renormalization of the one-particle/hole propagators.

\subsubsection{Transition-matrix elements}
\label{Peierls}
Concerning the transition matrix elements, we make use of the Peierls approximation \cite{millis_review,PhysRevB.67.115131,optic_prb,me_phd}.
Therein, Fermi velocities as the derivative of the bare dispersion%
\footnote{When the Hamiltonian of the system is expressed by (continuum) field operators, the charge density commutes with the interaction terms.} $v_{\svek{k}n}^{\alpha}=1/\hbar\partial_{k_\alpha}\epsilon^0_{\svek{k}n}$,
and
the matrix elements $M^{(B)}$ in \eref{eqL} and \eref{eqLB} can be expressed as \cite{Wenhu_thesis}
\begin{align}
M^{\alpha\beta}(\vek{k},n,n)&=v_{\svek{k}n}^{\alpha}v_{\svek{k}n}^{\beta} \label{eq:M}\\
M^{B,\alpha\beta\gamma}(\vek{k},n,n)&= \varepsilon_{\gamma i j} v^\alpha_{\svek{k}n} 
c^{\beta i}_{\svek{k}n} v^j_{\svek{k}n} 
\label{eq:Mb}
\end{align}
where $\varepsilon_{\gamma i j}$ is the Levi-Civita symbol and the curvature of the bare dispersion is encoded in
$c_{\svek{k}n}^{\alpha\beta}=1/\hbar\partial_{k_{\alpha}}\partial_{k_{\beta}}\epsilon^0_{\svek{k}n}$.
Standard Fermi velocities obtained in the band-basis only account for intra-band transitions. 
In a more general framework\cite{PhysRevB.45.13945,PhysRevB.67.115131} also inter-band transitions can be included in a Peierls-like fashion.

\section{Methodological Advancement}\label{sec:methadv}
\subsection{Approximation: Linearized Self-energy}
As seen in Eqs.~(\ref{K1}-\ref{K2}), 
the derivative of the Fermi function assures that
transport properties
are dominated by energies close to the Fermi level %
\footnote{This is contrary to thermodynamic properties, such as the specific heat, where all energy scales contribute and self-energy sum-rules have to be enforced.
}.
Then, also in the quantity that encodes many-body renormalizations---the electron self-energy $\Sigma$---only the low-energy behaviour is relevant.
Then, for the purpose of transport properties and in the absence of pole-like structures within several $k_B T$ of the  Fermi level, the self-energy can be linearized:
\begin{equation}
    \Sigma_{\mathbf{k}n}(\omega)\approx\Re\Sigma_{\mathbf{k}n}(0) + (1-Z_{\mathbf{k}n}^{-1})\omega - i \Gamma^0_{\mathbf{k}n}
    \label{eq:linSigma}
\end{equation}
In other words, the central assumption is that for transport properties the temperature dependence of renormalizations is more important than that on frequency. An implicit higher frequency dependence can, however, be included by linearizing the self-energy around the band-energies $\epsilon^0_{\mathbf{k}n}$. For the scattering rate, for instance, instead of evaluating $\Gamma^0_{\mathbf{k}n}=-\Im\Sigma_{\mathbf{k}n}(\omega=0)$ at the Fermi level, one can use $\Gamma^0_{\mathbf{k}n}=-\Im\Sigma_{\mathbf{k}n}(\omega=\epsilon^0_{\mathbf{k}n})$.

With \eref{eq:linSigma}, the coherent part of the spectrum (of weight $Z$) is of Lorentzian form:
\begin{equation}
    A_{\mathbf{k}n}(\omega)=\frac{Z_{\mathbf{k}n}}{\pi}\frac{\Gamma_{\mathbf{k}n}}{(\omega+\mu-\epsilon_{\mathbf{k}n})^2+\Gamma_{\mathbf{k}n}^2}
    \label{eq:lorentzian}
\end{equation}
with $\Gamma=Z\Gamma^0$ and $\epsilon=Z(\epsilon^0+\Re\Sigma(0))$, the renormalized scattering rate and dispersion, respectively.

\subsection{Linear Response Transport Quantities for finite Scattering}
\begin{figure}[!ht]
    \centering
    \includegraphics[width=0.45\textwidth, trim=0 0 0 -5mm]{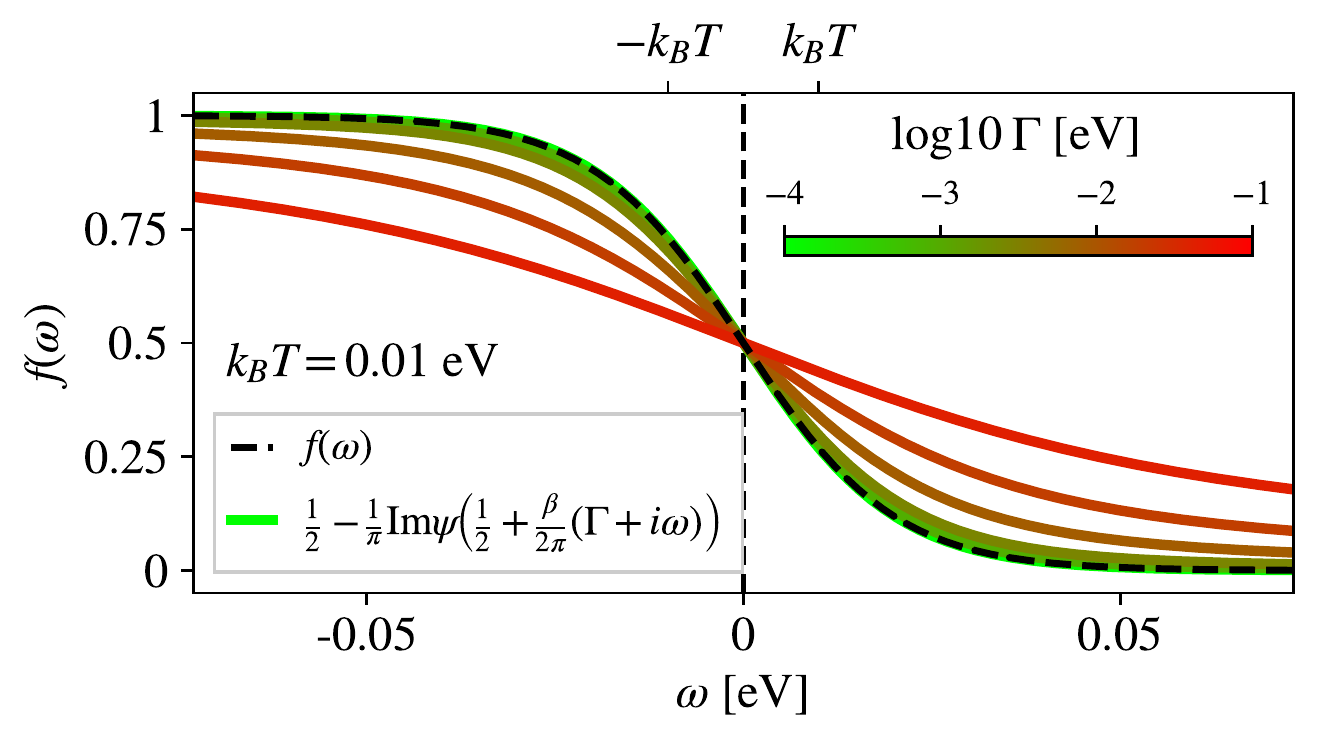}
    \caption{\textbf{Lifetime enhanced broadening.} Comparison between purely thermal broadening (dashed black line -- Fermi function $f(\omega)$) and lifetime enhanced broadening described by Eq.~(\ref{eq:digammocc}) for various scattering rates (solid colored lines).}
      \label{fig:fermidigam}
\end{figure}

The central innovation of this paper is the observation that, in the current setting, the integrals in Eqs.~(\ref{K1}-\ref{K2}) can be performed analytically---circumventing costly and (for small $\Gamma$) unstable numerical integrations.
Indeed, also the evaluation of the particle number simplifies, one finds\cite{jmt_fesb2}
\begin{equation}
    N=\sum_{\mathbf{k},n,\sigma}\int_{-\infty}^{\infty} d\omega f(\omega) A_{\mathbf{k}n}(\omega) = \sum_{\mathbf{k},n,\sigma}\left(\frac{1}{2} - \frac{1}{\pi} \Im \psi(z_{\mathbf{k}n}) \right)
\label{eq:digammocc}
\end{equation}
with the digamma function $\psi$ evaluated at $z_{\mathbf{k}n}=\frac{1}{2}+\frac{\beta}{2\pi}\left[\Gamma_{\mathbf{k}n}+ i(\epsilon_{\mathbf{k}n}-\mu)  \right]$, where $\beta=1/(k_BT)$ is the inverse temperature%
\footnote{In this expression, the quasi-particle weight $Z$ has been set to one, as, e.g., customary in slave-boson approaches. This procedure implicitly assumes the transfer of spectral weights, $1-Z$, to be symmetrical in the sense that it does not alter the chemical potential $\mu$. If a many-body electronic structure and, thus, $\mu$ is provided, the self-energy is only linearlized in the transport kernels.
}.
Finite lifetimes (inverse scattering rate) explicitly enter through the digamma function---describing the
thermal and lifetime smearing of excitations on an equal footing.
Consequently, the energy states now obey a $\Gamma$-modified Fermi-Dirac statistic, displayed in Fig.~\ref{fig:fermidigam}.
Crucially, even for $T=0$ this distribution is not step-like---provided that $\Gamma>0$.
In \sref{sec:parameterscan}, we explore the impact of the carrier density behaviour on the chemical potential and all derived transport properties.

For the intra-band transport kernels
$\mathcal{K}^{(B)}_{ab}(\mathbf{k},n)\equiv\mathcal{K}^{(B)}_{ab}(\mathbf{k},n,n)$ 
of Eqs.~(\ref{K1}-\ref{K2}) we derive the {\texttt{\textsc{LinReTraCe}}} expressions (see \aref{sec:app_kernels}):

\begin{widetext}
\begin{align}
&\mathcal{K}_{11}(\mathbf{k},n) = \frac{Z^2\beta}{4\pi^3\Gamma} \bigg[ \Re\psi_1(z) - \frac{\beta\Gamma}{2\pi} \Re\psi_2(z) \bigg] \label{LRT_k11}\\
&\mathcal{K}_{12}(\mathbf{k},n) = \frac{Z^2\beta}{4\pi^3\Gamma} \bigg[ a\Re\psi_1(z) - \frac{a\Gamma\beta}{2\pi} \Re\psi_2(z) -\frac{\Gamma^2\beta}{2\pi} \Im\psi_2(z) \bigg]\label{LRT_k12}\\
&\mathcal{K}_{22}(\mathbf{k},n) = \frac{Z^2\beta}{4\pi^3\Gamma} \bigg[(a^2+\Gamma^2) \Re\psi_1(z) +\frac{\beta}{2\pi}\Gamma\left(\Gamma^2-a^2\right) \Re\psi_2(z)  
- \frac{\beta}{\pi} a\Gamma^2\Im\psi_2(z) 
 \bigg]\label{LRT_k22}\\
&\mathcal{K}^B_{11}(\mathbf{k},n) = \frac{Z^3\beta}{16\pi^4\Gamma^2} \bigg[ 3 \Re\psi_1(z) - \frac{3\Gamma\beta}{2\pi} \Re\psi_2(z) + \frac{\Gamma^2\beta^2}{4\pi^2}\Re\psi_3(z) \bigg]\label{LRT_k11B}\\
&\mathcal{K}^B_{12}(\mathbf{k},n) = \frac{Z^3\beta}{16\pi^4\Gamma^2} \bigg[ 3a \Re\psi_1(z) - \frac{3a\Gamma\beta}{2\pi} \Re\psi_2(z) - \frac{\Gamma^2\beta}{2\pi} \Im\psi_2(z) + \frac{a\Gamma^2\beta^2}{4\pi^2}\Re\psi_3(z) + \frac{\Gamma^3\beta^2}{4\pi^2} \Im\psi_3(z)\bigg]\label{LRT_k12B}
\end{align}

\begin{align}
\mathcal{K}^B_{22}(\mathbf{k},n) = \frac{Z^3\beta}{16\pi^4\Gamma^2} \bigg[& (\Gamma^2+3a^2) \Re\psi_1(z)  - \frac{\beta\Gamma(\Gamma^2+3a^2)}{2\pi} \Re\psi_2(z) - \frac{\beta a\Gamma^2}{\pi} \Im\psi_2(z) \nonumber \\
&-\frac{\beta^2\Gamma^2(\Gamma^2-a^2)}{4\pi^2}\Re\psi_3(z) + \frac{\beta^2 a \Gamma^3}{2\pi^2} \Im\psi_3(z)  \bigg]\label{LRT_k22B}
\end{align}
\end{widetext}
where $\psi_i$ is the $i^{\hbox{\tiny th}}$-derivative of the digamma function $\psi$
evaluated at $z=\frac{1}{2} + \frac{\beta}{2\pi}(\Gamma+ia)$, with $a=\epsilon-\mu$.
Momentum $\mathbf{k}$ and band $n$ indices of $Z$, $\Gamma$, and $a$ have been omitted for "brevity".

The general feature of consistently treating the smearing of excitations due to thermal ($\beta$) and lifetime ($\Gamma$) effects 
{\it on an equal footing}---first noted in \eref{eq:digammocc}---also affects the transport kernels.

The above equations also allow for a simple symmetry analysis. For instance, we see that $\mathcal{K}_{12}$ is odd with respect to $a$. As a consequence, electron and hole contributions to the Seebeck coefficient, \eref{eq:ObservSeeb}, have opposite signs.
$\mathcal{K}_{11}^B$, instead, is even in $a$. Therefore, the Hall coefficient $R_H$, \eref{eq:ObservHallCoeff}, actually does not distinguish electron and hole contributions through their 
energies (the sign of $a$), but thanks to the sign of the dispersion's curvature entering the matrix element $M^B$ in \eref{eq:M}.

Manifestly, the above kernel functions are far more complicated than the familiar
expressions of the semi-classical Boltzmann approach in the constant relaxation time approximation
(cf., e.g., Refs.~\onlinecite{Madsen200667,MADSEN2018140,Pizzi2014422,Ponce_2020,PhysRevB.103.144404}).
However, the latter are recovered from the above formulae as the leading terms in the limit of infinite lifetimes, i.e.\ $\Gamma\rightarrow 0^+$.
This is most readily seen for the kernel underlying the conductivity: Noting that ${\beta}/{(2\pi^2)}\Re\psi_1[1/2+i\beta a/(2\pi)]=-f^\prime(a)$, one finds, to leading order in the scattering rate, the familiar expression 
\begin{equation}
\sigma_{\alpha\beta}=
\frac{e^2 Z^2}{V}\sum_{\svek{k}\sigma} \tau v_{\svek{k}}^\alpha v_{\svek{k}}^\beta \left( -\partial f/\partial\omega\right)_{\omega=\epsilon_{\svek{k}}-\mu},
\label{eq:boltzcond}
\end{equation}
with the lifetime $\tau=\hbar/(2\Gamma)$, and a renormalization factor $Z^2$ commonly not included.
From this point of view, the inclusion of finite lifetimes leads to 
\begin{enumerate}%[(a)]
\item An effectively different
statistic: the Fermi function is replaced with digamma functions in which thermal broadening is supplemented 
by an energy smearing $\Gamma$ corresponding to finite lifetimes (cf.\ \fref{fig:fermidigam}).
\item
 All transport kernels have, beyond the explicitly leading terms (e.g., $1/\Gamma$ in $\mathcal{K}_{11}$), contributions from higher powers in the scattering rate $\Gamma$. 
 \end{enumerate}
We will discuss the effects of both, later on.

Computational results in this paper have been obtained using the transport properties code
{\texttt{\textsc{LinReTraCe}}}\cite{LRT}.
While the Eqs.~(\ref{LRT_k11}-\ref{LRT_k12B}) are quite involved, they are, in fact, on par
with Boltzmann approaches employing the relaxation time approximation in terms of 
numerical complexity and evaluation speed. In fact, many Boltzmann codes\cite{Madsen200667,Boltztrap2,Pizzi2014422,EPW,Perturbo,TransOpt,protik2021elphbolt} could readily upgrade their electronic transport capabilities by switching to the kernels presented here.
Conversely, {\texttt{\textsc{LinReTraCe}}}\cite{LRT} could
benefit from being interfaced with code packages\cite{EPW,Perturbo,ShengBTE,protik2021elphbolt}
that provide electronic lifetimes from electron-phonon scattering.

\section{A New Phenomenology of Transport in Semiconductors}
\label{sec:results}
\subsection{Low-temperature expansion}
\label{sec:lowT}

Signatures of finite electronic lifetimes are most pronounced at low-temperatures, where qualitative deviations from Boltzmann behaviour are especially pronounced.
Our first goal therefore is to provide {\it simple} phenomenological formulae for transport observables at low temperatures.  To this end we first expand the polygamma functions
$\psi_i(1/2+\overline{z})$ in the kernel Eqs. (\ref{LRT_k11}-\ref{LRT_k12B})  around $\overline{z}=\infty$, i.e.\ $T=0$ (The resulting expressions can be found in \aref{sec:app_lowT}).
Second, we assume a simplistic electronic structure;
indeed, we note that in metals and semi-metals only states in the direct vicinity of the Fermi level contribute sizably to conduction of charge and heat. In gapped systems, instead, transport will be dominated by the conduction (valence) band minimum (maximum). 
Both constraints effectively limit the parts of the Brillouin zone relevant to transport.
To gain a qualitative insight, we therefore radically forgo the momentum integration in Eqs.~(\ref{eqL}-\ref{eqLB}): We consider a single non-dispersive level at an energy $\epsilon$ and constant transition matrix elements $M^{(B)}$. Assuming further a scattering rate $\Gamma$ independent of temperature, we obtain the following 
prototypical dependencies

\begin{widetext}
\begin{align}
\sigma&\propto 
e^2\frac{Z^2}{\pi^2} \frac{\Gamma^2}{(a^2+\Gamma^2)^2} \left[ 1 + \frac{2\pi^2}{3}\frac{5a^2-\Gamma^2}{(a^2+\Gamma^2)^2}\times (k_B T)^2 +\mathcal{O}(T^4)\right] \label{eq:zero1}\\
\sigma^B&\propto  
e^3\frac{4Z^3}{3\pi^3}\frac{\Gamma^3}{(a^2+\Gamma^2)^3}
\left[ 1 + \pi^2 \frac{7a^2-\Gamma^2}{(a^2+\Gamma^2)^2}
\times (k_B T)^2  +\mathcal{O}(T^4)\right] \label{eq:zero1B} \\
S&\propto - \frac{k_B}{e}  
\frac{4\pi^2}{3} 
\frac{a}{a^2 + \Gamma^2} \left[ k_B T +  \frac{\pi^2}{15} \frac{55a^2-53\Gamma^2}{(a^2+\Gamma^2)^2}   (k_B T)^3 + \mathcal{O}(T^5) \right] \label{eq:lowTS} \\
R_H&\propto 
\frac{4\pi}{3e}\frac{1}{Z\Gamma} \left[ a^2+\Gamma^2+ \frac{\pi^2}{3}(k_B T)^2+\mathcal{O}(T^4)  \right] \label{eq:lowTRH}\\
\kappa&\propto 
\frac{Z^2}{3}\frac{\Gamma^2}{(a^2+\Gamma^2)^2}  \left[k_B T- \frac{16 \pi^2 a^2}{3(a^2+\Gamma^2)^2} (k_BT)^3+\mathcal{O}(T^5) \right] \label{eq:kappalowT}\\
\nu&\propto 
- \frac{8\pi^2}{9}k_B\frac{a Z\Gamma}{\left(a^2+\Gamma^2\right)^2} \left[ k_BT -
\frac{4 \pi ^2}{3}\frac{ 8 a^2-\Gamma^2}{\left(a^2+\Gamma^2\right)^2} (k_BT)^3 +\mathcal{O}(T^5)\right] \label{eq:zero2}
\end{align}
\end{widetext}
 where $a=\epsilon-\mu$ indicates the position of the renormalized level $\epsilon=Z\epsilon^0$ with respect to the chemical potential $\mu$.
We now discuss the above asymptotic behavior and compare to Boltzmann approaches in the relaxation time approximation, see \tref{tab1} for a summary of the $T\to 0$ limit.
Note that the above equations describe the low-$T$ response for a single level. 
If several states are contributing, terms can be simply added up for the conductivities. For composite quantities, such as the Seebeck or the Hall coefficient, however, contributions to the Onsager coefficients, \eref{onsager}, have to be summed before they are combined into the observable quantities.

\begin{figure*}[!h!t]
    \includegraphics[width=0.95\textwidth]{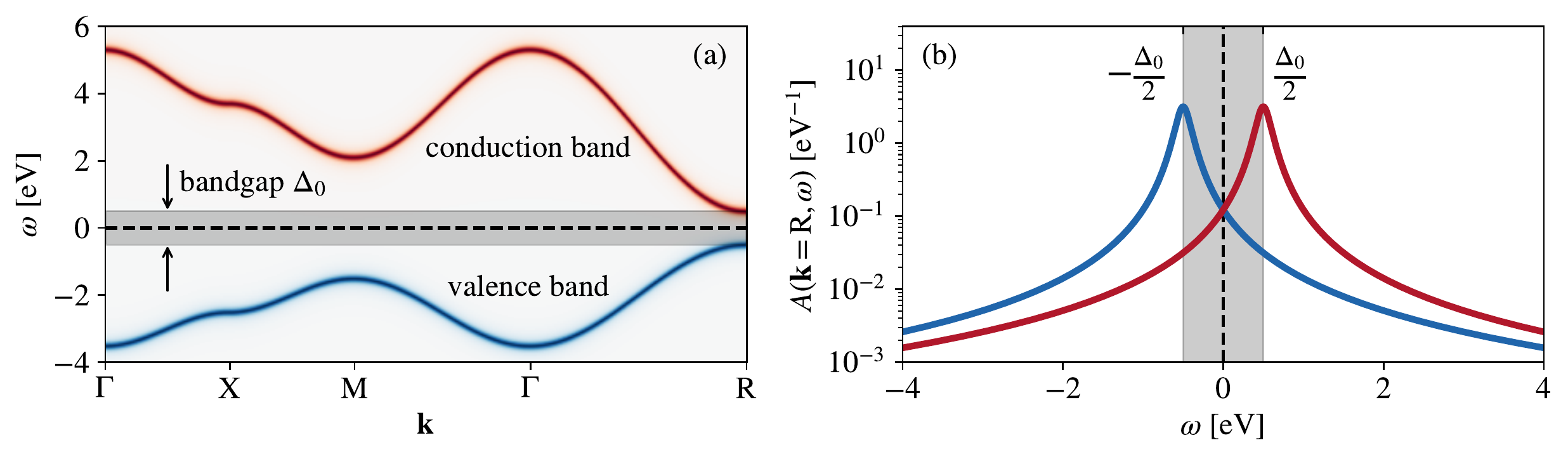}
    \caption{\textbf{Model electronic structure.} (a) Broadened band structure $\epsilon^0_{\mathbf{k}n}=-\sum_{i=x,y,z} 2t_n \cos\left(k_i\right) + (-1)^n (6t_n+\Delta_0/2)$ with $t_1 = 0.25$eV, $t_2 = -0.40$eV and $\Delta_0 = 1.0$eV. The broadening $\Gamma=0.1$eV is exaggerated in order to illustrate the effects. (b) Spectral functions at $\mathbf{k}=\mathrm{R}=\left(\pi,\pi,\pi\right)$ where the direct gap is exactly $\Delta_0$. The Lorentzian form, \eref{eq:lorentzian}, of the spectral functions causes weight to spill-over into the bandgap. As a consequence, for $T\to 0$, transport properties exhibit metal-like signatures, such as residual conduction (=resistivity saturation) and a linear-in-$T$ Seebeck coefficient.} 
    \label{fig:estruct}
\end{figure*}

\paragraph{Charge transport.}
One of the main observations is that in the zero temperature limit $T\rightarrow 0$ the electrical ($\sigma$)\cite{LRT_Tstar} and Hall conductivity ($\sigma^B$) remain finite in the presence of residual scattering ($\Gamma>0$). In Eqs.~(\ref{K1}-\ref{K2}) the derivative of the Fermi function becomes increasingly narrow with decreasing temperature. Alone, this  temperature-dependent energy cut-off would lead to the typical activated behavior and is well-described in Boltzmann theory.
However, for $\Gamma>0$, the Lorentzian shape of the spectral function, \eref{eq:lorentzian}, allows states away from the Fermi level to still contribute to conduction even at $T=0$, as incoherent spectral weight spills into the gap, see \fref{fig:estruct}(b) and cf.\ the effective particle distribution function in \fref{fig:fermidigam}.
This residual conductivity is at the heart of the resistivity saturation in (non-topological) Kondo insulators and has been discussed in detail in Ref.\ \onlinecite{LRT_Tstar}.
Similarly, the Hall conductivity in \eref{eq:ObservHallCond} and, hence, the Hall coefficient in \eref{eq:ObservHallCoeff} saturate for $\Gamma>0$.
The Boltzmann approximation, see \eref{eq:boltzcond}, on the other hand relies solely on the Fermi function to select states with sharply defined energies $\epsilon$. Then, conductivities must strictly approach zero in gapped systems for $T\to 0$.
Since the electrical and the Hall conductivity have the identical temperature scaling, \eref{eq:ObservHallCoeff} implies a diverging Hall coefficient in Boltzmann's relaxation time approximation, see \fref{fig:prototypical}(c)%
\footnote{
The higher order kernels $\KK_{12}^{(B)}$ and $\KK_{22}^{(B)}$ are accompanied by an additional $\omega$- and $\omega^2$-factor in Eqs.~(\ref{K1}-\ref{K2}), respectively. Therefore, the active energy window is additionally suppressed, causing these kernels to vanish for $T\to 0$. The exact temperature scaling is crucial (see Appendix \ref{sec:app_lowT}) since an additional $1 / T$-factor must be considered for some transport tensors in Eqs.~\ref{eq:ObservCond}-\ref{eq:ObservNernst}. 
Then, in our formalism, the thermal conductivity $\kappa$ and the Seebeck coefficient $S$ vanish for $T\to 0$.
}.

\paragraph{Thermoelectric transport.}
In essence the Seebeck and Nernst effect can be understood as entropy carried by charged currents\cite{zlatic,behnia,0034-4885-79-4-046502}.
The third law of thermodynamics states that at zero temperature the entropy $S_0$ of the system must be minimal. In a perfect crystal lattice without ground state degeneracy this minimal value must be $0$ since there is only one possible microstate ($S_0=k_B \ln(\Omega);\Omega=1$), requiring the Seebeck as well as the Nernst coefficient to vanish for $T\rightarrow 0$.
This is respected in our framework: Similar to the case of metals\cite{Behnia2004}, we find $S \sim T$ ($T\to 0$) in a semiconductor with finite lifetimes. As discussed in more detail in \sref{sec:MottS},
it is residual conduction from incoherent states that leads to a weakly metal-like Seebeck coefficient.
In the Boltzmann limit, instead,
$S$ unphysically diverges in a semiconductor: $S(T) \propto \frac{1}{T}$.

The Nernst coefficient vanishes in both formalisms. 
In the Boltzmann case, this is hidden in the two terms making up \eref{eq:ObservNernst} ($\nu = \nu_1 - \nu_2$): While both $\nu_i$ ($i=1,2$) diverge, they cancel exactly when combined.
In the Kubo formalism, both terms $\nu_i$ separately approach $0$. Further, for finite $\Gamma$, $\nu\propto T$ at lowest temperatures---again akin to the behaviour of metals\cite{0953-8984-21-11-113101} and (see \sref{sec:Mottnu}) connected to conduction from intrinsic, but incoherent in-gap states.

In all, Eqs.~(\ref{eq:zero1}-\ref{eq:zero2}) establish a low-temperature phenomenology of transport in semiconductors. The derived asymptotic behavior 
overcomes limitations of semi-classical descriptions and is congruent with experimental observations (see \fref{fig:prototypical} above and \fref{fig:materials} below).

\begin{table}[!h]
\begin{tabular}{l|c|c}
 $\lim T\to 0$    & \verb=LinReTraCe= & Boltzmann      \\\hline
$\rho$  & $\rho_{\mathrm{sat}}$  &  $\infty$  \\
$S$    & $0$    & $\infty$     \\
$\kappa$ & $0$ & 0 \\
$R_H$ & $R_{H,\mathrm{sat}}$  &  $\infty$     \\
$\nu=\nu_1-\nu_2$   & $0$       & $0$ \\
$\nu_{1/2}$ & $0$ & $\infty$
\end{tabular}
\caption{{\bf Zero temperature limits of transport properties in stoichiometric gapped systems 
for a finite scattering rate $\Gamma$.} Eqs.~(\ref{eq:zero1}, \ref{eq:lowTRH}) lead to
saturation in the resistivity and the Hall coefficient, while the Boltzmann signal diverges. Entropy transport complies with the laws of thermodynamics (Seebeck $S\to 0$ for $T\to 0$), while $S$ unphysically diverges in the relaxation time approximation. 
$\nu_{1/2}$ denote the two contributions to the Nernst coefficient, \eref{eq:ObservNernst}.
All limiting behaviors of \texttt{LinReTraCe}
are congruent with experiments, see \fref{fig:prototypical} and \fref{fig:materials}.
\label{tab1}}
\end{table}

\subsection{Prototypical transport properties of narrow-gap systems}
\label{sec:parameterscan}
In this section we leave the low temperature limit and study the full temperature dependence of the transport observables from Eqs.~(\ref{eq:ObservCond}-\ref{eq:ObservNernst}). Motivated by experimental transport measurements of intermetallic hybridization-gap semiconductors---such as FeSb$_2$, FeAs$_2$, FeSi, FeGa$_3$, their Ru-analogues and others---we consider
a simple, asymmetric two-band electronic structure
\begin{equation}
    \epsilon^0_{\mathbf{k}n}=-\sum_{i=x,y,z} 2t_n \cos\left(k_i\right) + (-1)^n (6t_n+\Delta_0/2)
    \label{eq:electrstruct}
\end{equation}
with $n=1,2$ for the valence and conduction band, respectively,
and fix the filling to $N=2$ (half-filling).
We use a generic lattice constant $a_{\mathrm{lattice}}=1$\AA\ (cf.\ \aref{sec:app_latticescaling}) and $60\times 60\times 60$ ($200 \times 200 \times 200$) $k$-points for the Kubo (Boltzmann) calculations to achieve k-grid convergence. Particle-hole asymmetry is introduced
by hopping parameters, $t_n$, that are different for the valence band (VB; $n=1$) and the conduction band (CB; $n=2$). We measure the degree of asymmetry via
\begin{equation}
    \alpha = \left\vert \frac{t_2}{t_1} \right\vert.
\end{equation}
The two bands of the narrow-gap semiconductor are then additionally endowed with the same, finite and---for the time being---temperature-independent scattering rate $\Gamma$, while we set the quasi-particle weight to unity, $Z=1$.
An example electronic structure is displayed in \fref{fig:estruct}.
Even though motivated by said materials,
we keep the electronic structure deliberately simple, so as to isolate qualitative trends and the prototypical temperature dependencies of transport properties.
A more realistic setup with a temperature-dependent scattering rate will be introduced in \sref{sec:tempgamma}.

\begin{figure*}[!t!h]
    \includegraphics[width=0.98\textwidth]{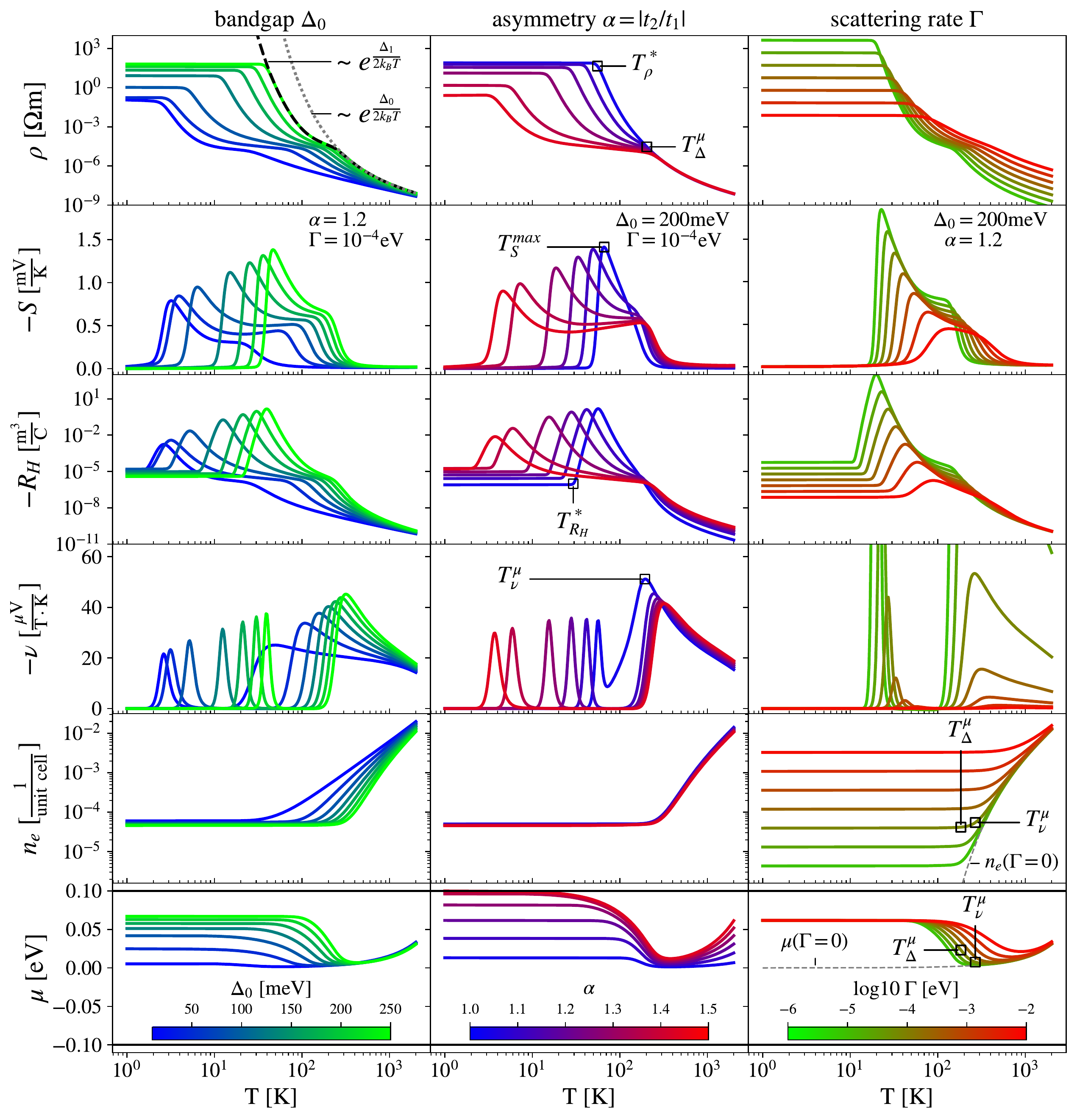}\\
    \caption{\textbf{Transport in semiconductors: A scan of parameter.} Left column: varying bandgaps $\Delta_0$ for fixed hoppings $t_1=0.25$eV and $t_2=-0.30$eV (asymmetry $\alpha=1.2$) and scattering rate $\Gamma=10^{-4}$eV -- Middle column: varying asymmetries $\alpha$ for fixed bandgap $\Delta_0=200$meV and scattering rate $\Gamma=10^{-4}$eV -- Right column: varying scattering rates $\Gamma$ for fixed bandgap $\Delta_0=200$meV and asymmetry $\alpha=1.2$. Throughout no quasi-particle renormalization is used ($Z=1$).
    From top to bottom we show the resistivity $\rho$, Seebeck coefficient $S$, Hall coefficient $R_H$, Nernst coefficient $\nu$, the activated number of electrons $n_e$ (holes $n_h=n_e$), and the chemical potential $\mu$.
    Notable characteristics: Due to the conduction and valence band asymmetry the chemical potential does not converge to the gap mid-point for $T\to 0$. This deviation causes the high temperature features at $T_{\Delta}^{\mu}$ and $T_{\nu}^{\mu}$. There, the resistivity transitions between an activated regime corresponding to the fundamental gap $\Delta_0$ to one with a reduced energy $\Delta_1$. The characteristic behaviors at lower temperatures $T_S^{\max}$, $T_{\rho}^*$ and $T_{R_H}^*$ are driven by the quantum kernels and the importance of higher order polygamma functions in them: The resistivity and the Hall coefficient saturate at $T_{\rho}^*$ and $T_{R_H}^*$, respectively. The latter signals a finite density of carriers at the Fermi level even at absolute zero, congruent with $n_e>0$.
    The Seebeck coefficient is suppressed and vanishes for $T\to 0$ in accordance with the laws of thermodynamics. The Nernst coefficient peaks one more time before also trending to zero for $T\to 0$. 
 }
    \label{fig:parameterscan}
\end{figure*}

The three parameters that describe the transport equations of our model are the bandgap $\Delta_0$, the band asymmetry $\alpha$ and the scattering rate $\Gamma$.
As a first step we simply scan through each parameter individually while keeping the other two fixed.
 The results for these parameter sweeps are shown in \fref{fig:parameterscan}.
 The clear protagonist of transport properties beyond  Boltzmann results based on band structures is the scattering rate $\Gamma$. It influences transport in two ways: First, $\Gamma>0$ leads---in a particle-hole asymmetric system---to a non-trivial temperature dependence of the chemical potential, which, in turn, influences charge and energy transport.
 Boltzmann approaches that use band structures as input fully miss this ingredient as only thermal (not lifetime) broadening is included in the chemical potential search.
 Second, contributions to the  
 transport kernels, Eqs.\ (\ref{LRT_k11}-\ref{LRT_k12B}), are---contrary to Boltzmann approaches in the relaxation-time approximation---not simply proportional to the carrier lifetime, $1/\Gamma$, but exhibit an intricate $\Gamma$-dependence that influences the temperature profile of transport properties. In the following, we will disentangle these two ingredients.

\subsubsection{Chemical potential and activated carriers}
For sharply defined valence and conduction states $\epsilon_{\svek{k}}$, i.e.\ $\Gamma=0$, the position of the chemical potential $\mu$ is driven through the thermal broadening of the Fermi function. In particular, one can show (see, e.g., Ref.\ \onlinecite{jmt_fesb2}) that $\mu$
approaches the middle of the gap for $T \to 0$, with a temperature slope that depends on the particle-hole asymmetry. 
If the lifetime of valence and conduction states is finite, this description is no longer valid.
Indeed, determined via \eref{eq:digammocc}, the chemical potential is intrinsically dependent on the scattering rate. 
\fref{fig:estruct}(b), that displays a spectral function at a selected $k$-point, illustrates why this is the case: The Lorentzian width of the spectral function results in small but finite weight of incoherent in-gap states that the chemical potential has to account for.
As seen in the bottom panel of \fref{fig:parameterscan}, 
$\mu$ follows the result of the Fermi function at high temperatures,
where thermal broadening dominates over the Lorentzian in-gap weight.
Below a temperature $T_{\nu}^{\mu}$, however,
$\mu$ starts to strongly deviate. In particular, it no longer extrapolates to the midgap point (here set to zero), but to a  finite value that increases with growing particle-hole asymmetry $\alpha$ and bandgap $\Delta_0$, while being only weakly dependent on the residual scattering $\Gamma$.
This behavior can be rationalized through
the low temperature expansion of the occupation in Eq.~\ref{eq:digammocc}, which, for a state at $a=\epsilon-\mu$, yields
\begin{equation}
    N = \frac{1}{2} - \frac{1}{\pi}\Im\ln(\Gamma+ia) +\frac{\pi}{3}\frac{a\Gamma}{(a^2+\Gamma^2)^2} (k_B T)^2 + \mathcal{O}(T^4).
\end{equation}
 Given that, in a semiconductor, the scattering rate $\Gamma$ is orders of magnitude smaller than the band energies, changes in the band structure ($a=\epsilon-\mu$) will dominate the chemical potential at low temperatures.
 Noteworthy, the evolution of the chemical potential $\mu$ shown in the lower panels of \fref{fig:parameterscan}---its deviation from the intrinsic Fermi-Dirac result (gray dashed in bottom right panel) at $T_{\nu}^{\mu}$, its inflection point $T_\Delta^\mu$, and the eventual saturation at a finite position---is reminiscent of the exhaustion and extrinsic regime
 in semiconductors with impurity-derived in-gap states\cite{ziman}. 
 There, changes in the chemical potential are driven by donated electrons or holes, i.e.\ a change in the total number of charge carriers.
In our scenario instead, the total number of electrons stays constant, but the finite lifetimes of intrinsic carriers causes excitations to widen, spilling incoherent spectral weight into the gap, so that the chemical potential has to adapt.
Consequently, even at lowest temperatures, the number of activated carriers 
 \begin{align}
     n_e &= \sum_{\mathbf{k},n\geq\mathrm{CB},\sigma} \left(\frac{1}{2} - \frac{1}{\pi}\Im\psi(z_{\mathbf{k}n)}\right) \\
     n_h &= \sum_{\mathbf{k},n\leq\mathrm{VB},\sigma} \left(\frac{1}{2} + \frac{1}{\pi}\Im\psi(z_{\mathbf{k}n)}\right)
 \end{align}
 \emph{must} remain {finite} for $\Gamma>0$, as shown in 
 the Fig.~\ref{fig:parameterscan} (second panel row from the bottom).
 In more detail, at any temperature,
 the number of activated electrons $n_e$ and holes $n_h$  (per unit-cell)
necessarily balance each other, $n_e=n_h$, in the stoichiometric (half-filled) case considered here.
 At high temperatures $n_e$  follows the result for the coherent ($\Gamma=0$) band structure (indicated in dashed gray). 
 In this regime, dominated by thermal activation across the gap $\Delta_0$, $n_e$ is exponentially suppressed upon cooling; for $k_BT\ll\Delta_0$: $n_e(\Gamma=0;T) \sim e^{-{\Delta_0}/{k_BT}}$.
 For finite $\Gamma$, the discussed deviations in the chemical potential reflect in the carriers available for conduction: At the temperature $T_{\nu}^{\mu}$, $n_e$ no longer shrinks exponentially and, at $T_{\Delta}^{\mu}$, transitions into a regime in which the number of available carriers is virtually independent of temperature. In this low-temperature regime the dominant control parameter for the number of carriers is the scattering rate $\Gamma$, whereas asymmetry and the size of the gap hardly affect $n_e(T\to 0)$ on the shown exponential scale.

\subsubsection{Electric resistivity}
Signatures of the described evolution of 
the number of carriers available for conduction are 
 readily seen in the resistivity in the top row of \fref{fig:parameterscan}. The activated behavior above $T_\nu^\mu$---purely determined by the bandgap $\Delta_0$---transitions into a second activated regime realized below $T_\Delta^\mu$, with an effectively reduced bandgap $\Delta_1 < \Delta_0$. 
 Again, this behavior is reminiscent of an impurity-driven extrinsic regime.
 There, $\Delta_1$ would measure the difference between the extrinsic impurity level on the one hand and the conduction or valence band on the other. 
 In both scenarios---extrinsic in-gap states vs.\ intrinsic states with finite lifetimes---changes in conduction reflect a modification in the chemical potential. Here, our theory provides a complementary microscopic origin for the appearance of the chemical potential-driven characteristic temperature scales $T_{\nu}^{\mu}$ and $T_{\Delta}^{\mu}$.
 Note that for particle-hole symmetric systems, where the chemical potential is temperature independent, no such crossover exists and there is only one activation-like regime\cite{LRT_Tstar}.

 Cooling further, also the second activated regime is bounded from below: At a temperature $T_\rho^*$, the resistivity enters a saturation regime. Contrary to the features at higher temperatures, $T_\rho^*$ has no signature in the chemical potential, but derives entirely from the physics encoded in the kernel function \eref{LRT_k11}.
As discussed in detail in Ref.~\onlinecite{LRT_Tstar} the crossover temperature $T_\rho^*$ and the saturation value $\rho(T\to 0)$ strongly depend on the scattering rate and the bandgap. Noteworthy, the influence of $\Gamma$ inverts as a function of temperature: At high $T$,  a larger scattering rate increases the resistivity. This is the conventional behavior, also realized in metals. At low $T$, however, where conduction is driven by incoherent spectral weight inside the gap, the resistivity understandably {\it decreases} with a growing scattering rate (see top right panel in \fref{fig:parameterscan}).
Here, we extend the previous analysis\cite{LRT_Tstar} and demonstrate that also the band asymmetry $\alpha$ has a strong effect on the conduction. In the asymmetric case, the chemical potential must be positioned closer to the conduction ($\alpha>1$) or valence ($\alpha<1$) band so that the correct number of electrons in the system is occupied. Therewith, the majority of carriers---those that reside in the centres of the Lorentz-broadened peaks in the spectral function---conduct more and freeze out at a lower temperature, i.e.\ $T_\rho^*$ and the corresponding saturation value $\rho(T\rightarrow 0)$ decreases with $\alpha$.

To summarize, the resistivity of an {\it intrinsic} narrow-gap semiconductor with a finite scattering rate has four regimes: (1) $T>T_{\nu}^{\mu}$: the activated high-temperature region that is well-described in Boltzmann theory; 
(2) $T_{\Delta}^{\mu}<T<T_{\nu}^{\mu}$: a narrow regime in which the chemical potential starts to sense the incoherent spectral weight inside the gap and adjusts accordingly;
(3) $T_\rho^*<T<T_{\Delta}^{\mu}$: a regime at intermediate temperatures in which the ($\Gamma$-imposed) chemical potential shift has led to a reduced activation energy for valence ($\alpha<1$) or conduction ($\alpha>1$) carriers; (4) $T<T_\rho^*$: a regime of resistivity saturation in which thermal activation is frozen out but a residual conductivity, driven by incoherent in-gap weight, remains finite.

\subsubsection{Seebeck coefficient}
The three temperatures, 
$T_\rho^* < T_{\Delta}^{\mu} < T_{\nu}^{\mu}$, 
that separate the four regimes in the electrical resistivity also account for features in the Seebeck coefficient (second row in \fref{fig:parameterscan}): The increase of $S$ starting from high temperatures is interrupted by the crossover of the chemical potential at $T_{\nu}^{\mu}$. Depending on the parameters, the transition to the maximum amplitude at lower temperatures can then either be smooth (large scattering rate) and monotonous (large gap) or be accompanied by a significant shoulder (large bandgap, small scattering rates). In extreme cases this shoulder transforms into a local peak (small gap, strong asymmetry), i.e.\ the temperature dependence is non-monotonous. The temperature at which the Seebeck coefficient has its global maximum amplitude $S^{\max}$ is linked to $T_\rho^*$. In fact, $S$ peaks at $T_S^{\max}$, consistently slightly above the onset of the resistivity saturation regime ($T_S^{\max}\gtrsim T_\rho^*$).
Below this global peak temperature, the Seebeck coefficient drops rather abruptly. In the zero temperature limit, it follows the metal-like linear behavior $S(T)\sim T$, anticipated in \sref{sec:lowT}.

This rich structure is {\it absent} when the Boltzmann approach is applied to the band structure $\epsilon_{\mathbf{k}n}^0$ of \eref{eq:electrstruct}:
The features associated with $T_{\Delta}^{\mu}$ and  $T_{\nu}^{\mu}$
are missed if finite lifetimes are unaccounted for in the search of the chemical potential; the characteristic features further below are absent owing to the simple structure of the Boltzmann transport kernels. Indeed, for a momentum- and state-independent scattering rate $\Gamma$, the kernels $\mathcal{L}_{11}$ and $\mathcal{L}_{12}$ in the Boltzmann approximation are both merely proportional to $\Gamma^{-1}$. Then, given by their ratio, \eref{eq:ObservSeeb}, Boltzmann's Seebeck coefficient is independent of the scattering rate. Manifestly, this approximation is a severe oversimplification even for extremely small $\Gamma$ (see \fref{fig:parameterscan}: right column, second panel from the top).

\begin{figure}[!h!t]
    \includegraphics[width=0.45\textwidth]{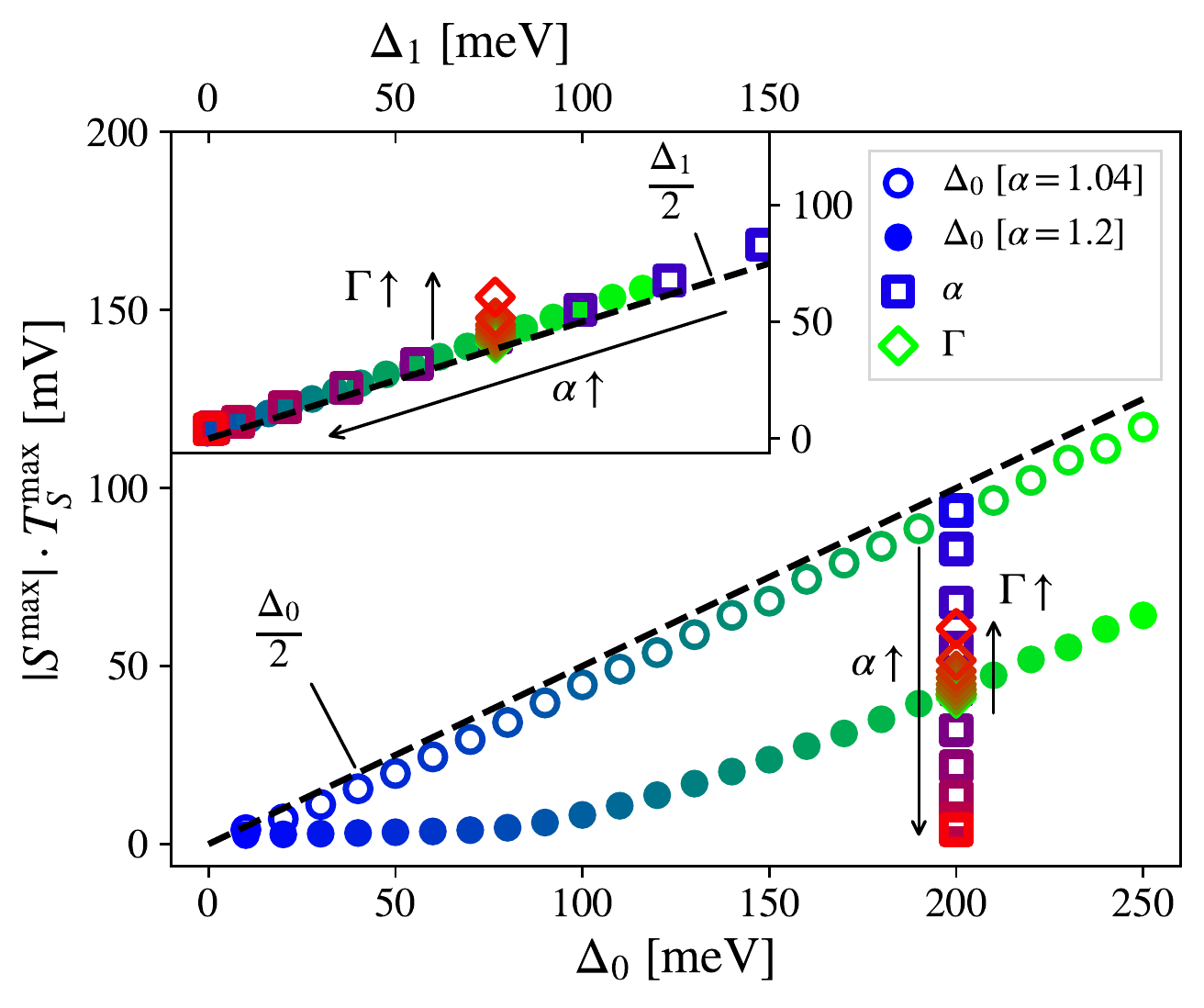}
    \caption{\textbf{Goldsmid-Sharp gap.} 
    We display the gap-estimate $|S^{\max}|\cdot T_S^{\max}$ for scans of the bandgap $\Delta_0$ for asymmetries $\alpha=1.04$ (open circles) and $\alpha=1.2$ (closed circles). Also shown is a scan of the  asymmetry (open squares) and of the scattering rate (open diamonds), using the same parameters as in Fig.~\ref{fig:parameterscan}.
    The expression $|S^{\max}| \cdot T_S^{\max}$ deviates strongly from ${\Delta_0}/{2}$ (dashed line) when the system's asymmetry is at least moderate.  Only in the vicinity of particle-hole symmetry ($\alpha=1.04$, open circles) do we find good agreement for \eref{eqGS}. An increase in asymmetry leads to a suppression of $|S^{\max}| \cdot T_{\max}$ while different scattering rates have minimal effects.
    Instead, plotting $|S^{\max}|\cdot T_{\max}$ against the
    effective gap $\Delta_1$ (cf.\ $\rho(T)$ in \fref{fig:parameterscan}), the various scans collapse onto the ${\Delta_1}/{2}$ line (see inset): The Goldsmid-Sharp gap expression reliably estimates the effective gap $\Delta_1$.
    }
    \label{fig:smaxtmax}
\end{figure}

Next, we will comment on two approximate tools that are popular for the analysis of thermoelectric measurements or simulations.
\paragraph{The Goldsmid-Sharp gap.}
Goldsmid and Sharp\cite{goldsmid} motivated that the size of a semiconductor's
gap could be gauged from the peak amplitude of the Seebeck coefficient:
\begin{equation}
\Delta\approx 2 e
|S^{\max}|\cdot T_S^{\max}.
\label{eqGS}
\end{equation}
This estimate works decently for both $n-$ and $p$-type semiconductors\cite{goldsmid}, although deviations of the order of a factor of two are not uncommon. The relation is used as a coarse analysis tool\cite{https://doi.org/10.1002/advs.201902409,Kutepov2020c,PhysRevB.103.085202}
in simulations and experiments and even as a descriptor in high-throughput materials discovery studies\cite{D1MH00751C}.
\eref{eqGS} was deduced for a coherent, large gap, particle-hole symmetric semiconductor
in which impurity states move the chemical potential so as to optimize the Seebeck coefficient\cite{goldsmid,jmt_fesb2}.
Allowing for particle-hole asymmetry, the Seebeck coefficient can, however, be larger,
while it is always bounded by $|S|\leq |\Delta/(eT)+S(\infty)|$, where $S(\infty)$ is the comparatively small high temperature limit ($|S(\infty)|=\mathcal{O}({k_B/e})$)\cite{jmt_fesb2}.
Since the original argument\cite{goldsmid} relies on replacing the Fermi-Dirac with the classical Maxwell-Boltzmann statistics,
further deviations occur if $k_BT_S^{\max}\not\ll\Delta$\cite{doi:10.1063/1.4905922}.

Here, we scrutinize the Goldsmid-Sharp relation, \eref{eqGS}, for our two-band model:
In \fref{fig:smaxtmax}
we report $|S^{\max}|\cdot T_S^{\max}$ extracted from the parameter scans of \fref{fig:parameterscan} as well as for an additional scan for an almost particle-hole symmetric system, $\alpha=1.04$ (open circles in the main panel).
For the latter, the Goldsmid-Sharp gap indeed provides a very accurate estimation of the fundamental gap $\Delta_0$.
For systems with more asymmetrical electronic structures, however, \eref{eqGS} yields poor results: $|S^{\max}|\cdot  T_S^{\max}$ largely underestimates the charge gap (filled circles).
Looking at the corresponding resistivities, Seebeck coefficients and the chemical potential in \fref{fig:parameterscan} reveals the reason:
The thermopower is largest at the lower end of the second activated regime of the resistivity, $T_\rho^*\lesssim T_S^{\max}<T_{\Delta}^{\mu}$. This regime emerges when a finite scattering rate pushes the chemical potential towards the lighter band (in our case the conduction band).
As a consequence, the crucial gap, as apparent in the resistivity, is the effective $\Delta_1$, not $\Delta_0$.
This observation suggests that the Goldsmid-Sharp gap does not monitor the bandgap $\Delta_0$ but the effective gap $\Delta_1$. This hypothesis is confirmed in the inset of \fref{fig:smaxtmax}: plotted as a function of the emergent $\Delta_1$, the scatter plot of $|S^{\max}|\cdot T_S^{\max}$ collapses onto a single line indicating $\Delta_1/2$.
This statement is largely independent of the scattering rate: $|S^{\max}|\cdot T_S^{\max}$ only slightly increases with $\Gamma$,
leading to a mild overestimation of $\Delta_1$.

In all, in the realistic presence of a scattering rate, the Goldsmid-Sharp expression quite accurately gauges the effective gap $\Delta_1$ that controls transport at intermediate temperatures---but which can be significantly smaller than the true bandgap $\Delta_0$.

\begin{figure*}[!h!t]
    \includegraphics[width=0.95\textwidth]{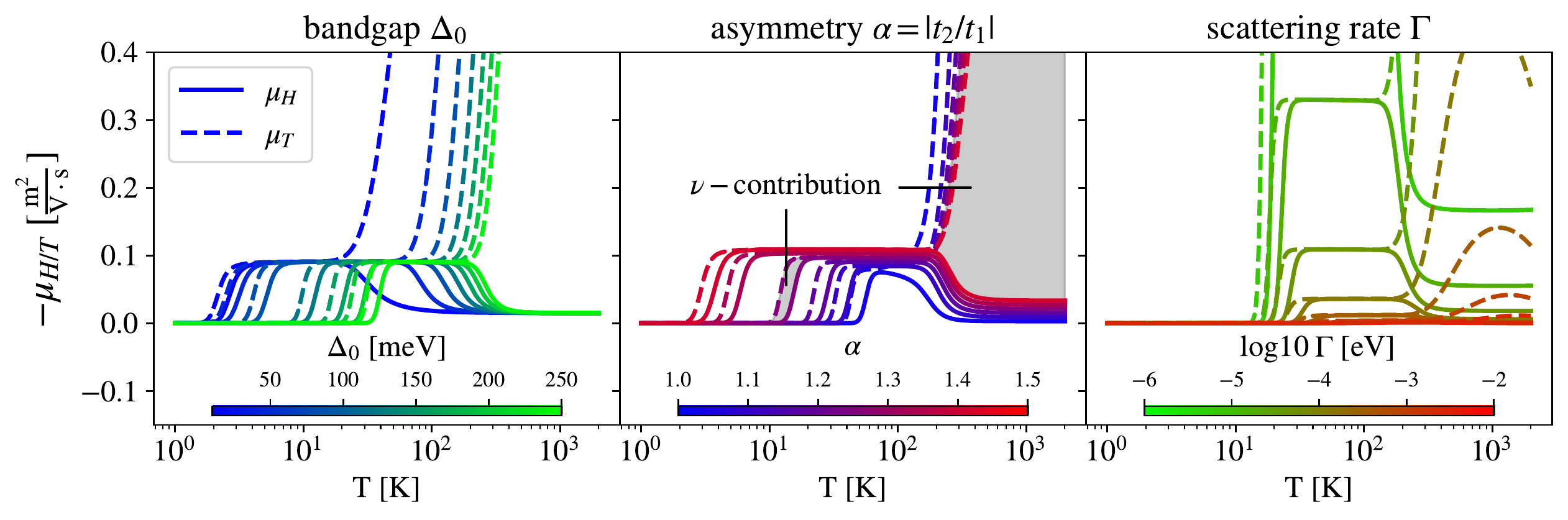}
    \caption{\textbf{Hall and thermal mobilities.} Both mobilities vanish in the zero temperature limit and coincide with each other at intermediate temperatures, marking the ranges where the Nernst coefficient vanishes in Fig.~\ref{fig:parameterscan}. This transition however takes place at slightly different temperature between $\mu_H$ and $\mu_T$ giving rise to the low temperature $\nu$-contribution at roughly $T_\rho *$ (first shaded area). At large temperature the two mobilities diverge from each other. $\mu_T$ increases drastically while $\mu_H$ gets suppressed, marking the second $\nu$-contribution which is peaked at $T_{\nu}^{\mu}$ (second shaded area).
    }
    \label{fig:mobilities}
\end{figure*}

\paragraph{The Mott formula.}\label{sec:MottS}
In metals, conduction is largely dominated by states in the vicinity of the chemical potential.
Then, performing a low temperature Sommerfeld expansion of the 
conductivity is justified. Doing so for the Boltzmann relaxation-time approximation, \eref{eq:boltzcond},
yields a
convenient expression for the Seebeck coefficient
\begin{equation}
 S \approx - \frac{\pi^2}{3e} k_B^2 T \frac{\partial \ln\sigma(\mu)}{\partial\mu},
\label{eq:MottS}
\end{equation}
which is a simplified version of the so-called {\it Mott formula} of the thermopower\cite{PhysRev.181.1336,PhysRevB.21.4223}. Here, $\sigma(\mu)$ is the electrical conductivity for varying chemical potential $\mu$.
Clearly, the above approximation is inaccurate for {\it coherent} semiconductors, where conduction is driven by conduction and/or valence states that are far (more than several $k_BT$) from the chemical potential. Manifestly, our general kernel functions therefore do not verify \eref{eq:MottS}.
However, as we demonstrated\cite{LRT_Tstar}, finite lifetimes may drive residual conduction in semiconductors, leading to resistivity saturation.
The incoherent in-gap states associated with this phenomenon might provide the metalicity required to  justify expanding the derivative of the Fermi function around the chemical potential (Sommerfeld expansion). 
Therefore, \eref{eq:MottS} is expected to hold in the saturation regime, where conduction is dominated by said incoherent in-gap weight.
Evaluating \eref{eq:MottS} for the residual term of the conductivity given in \eref{eq:zero1},
indeed yields the lowest-order expression of the low-T Seebeck coefficient, \eref{eq:lowTS}. This validity of the Mott formula for metals in the resistivity saturation regime of a semiconductor 
provides
a direct link between residual charge conduction ($\sigma(T\to 0)>0$) and a metal-like linear-in-$T$ thermoelectric Seebeck signal ($S(T\to 0)\propto T$).

\subsubsection{Hall coefficient}
From high temperatures down to its maximum, the Hall coefficient $R_H$ (third panel in \fref{fig:parameterscan}) exhibits a qualitatively similar dependency on $\Delta_0$, $\alpha$, and $\Gamma$ as the Seebeck coefficient. However, $R_H$ peaks at a slightly smaller temperature that matches $T_\rho^*$ from the resistivity. 
Also, instead of vanishing, $R_H$
saturates below a temperature $T_{R_H}^*$ ($<T_{\rho}^*$), in agreement with the low-$T$ expansion \eref{eq:lowTRH} and experiment, see \fref{fig:prototypical} and \fref{fig:materials}. This temperature arrangement is expected since the spectral function enters Eq.~(\ref{K2}) to higher order compared to Eq.~(\ref{K1}). 
Therefore, the $\mathcal{K}_{11}^B$ kernel effectively senses a smaller amount of incoherent in-gap weight than $\mathcal{K}_{11}$. Consequently,
the temperature, below which the thermal selection of valence and conduction carriers via $\left(-\frac{\partial f}{\partial\omega}\right)$ can be neglected in comparison to the incoherent in-gap weight that drives the residual conduction, is effectively reduced.
At large temperatures ($T>T_{\nu}^{\mu}$, Boltzmann regime), $R_H$ is dominantly controlled by the bandgap and the particle-hole asymmetry, while the dependence on the scattering rate is weak. Instead, at low $T$, it is mostly the scattering rate that controls, both, the peak value and the saturation limit $R_H(T\rightarrow 0)$.

\subsubsection{Nernst coefficient}
The Nernst coefficient exhibits the most dramatic features%
 \footnote{Our survey of the Nernst coefficient can be compared to the semi-classical relaxation-time approximation pioneered recently in Ref.~\cite{PhysRevB.103.144404} that includes effects to leading order in $1/\Gamma$.}.
Starting from high temperature, $\nu$ increases and reaches a first peak at $T_{\nu}^{\mu}$ where the chemical potential starts  to transition towards its saturation regime.
The Nernst then quickly drops to zero (unless the asymmetry is very small) and remains suppressed in the intermediate regime ($T_\rho^*<T<T_{\Delta}^{\mu}$; the second activation regime of $\rho$). The transition into the $\rho$-saturation regime at $T_\rho^*$ is then accompanied by a second large and sharp peak in $\nu$ before it finally converges linearly to zero for $T\to 0$. 
 Similar to the Hall coefficient, the biggest changes in the temperature profile of $\nu$ are achieved by varying the bandgap and the asymmetry, while absolute values are mostly controlled by the scattering rate.

Focusing on (a) the large low-temperature peak and (b) the $T\to 0$ limit, we analyse the fabric of the Nernst coefficient through different representations.

\paragraph{Hall and thermal mobilities.}
Using the expression \eref{eq:ObservSeeb} of the Seebeck coefficient, we can rewrite the Nernst coefficient \eref{eq:ObservNernst} as
\begin{equation}
    \nu=S\left( \mu_H - \mu_T \right)
    \label{eq:numobilities}
\end{equation}
where $\mu_H=\mathcal{L}_{11}^B/\mathcal{L}_{11}$ is the Hall mobility, and $\mu_T=\mathcal{L}_{12}^B/\mathcal{L}_{12}$ its thermal analogue, introduced by Sun \etal\cite{PhysRevB.88.245203}.
From this point of view, a finite Nernst signal can only appear when there is a mismatch in the mobilities.
The two mobilities, $\mu_H$ and $\mu_T$, are shown in \fref{fig:mobilities} for varying (a) bandgap, (b) asymmetry, and (c) scattering rate. 
At large $T$, $-\mu_T\gg -\mu_H$, resulting in a sizeable Nernst coefficient. At intermediate temperatures, both mobilities exhibit a temperature- and gap-independent plateau of {\it equal} magnitude: The Nernst signal vanishes.
At low $T$ there is again a mobility mismatch, $-\mu_T>-\mu_H$, which is responsible for the pronounced low temperature peak.
Qualitatively, this behavior mirrors the analysis of the experimental mobilities of FeSb$_2$ from Ref.~\cite{PhysRevB.88.245203} that we reproduce in the inset of \fref{fig:materials}(h): Also in experiment, the mobility mismatch is sizable at low and high temperatures, while in between they almost match.

\paragraph{Mott formula for the Nernst coefficient.}\label{sec:Mottnu}
Analogous to the Mott formula of the Seebeck coefficient, \eref{eq:MottS}, a Sommerfeld expansion can be used to obtain an approximate formula for the Nernst coefficient.
Using $\mathcal{L}_{12}^B \approx  \frac{\pi^2k_B^2}{3e} T^2 \frac{\partial \mathcal{L}_{11}^B(\mu)}{\partial\mu}$,  
valid at low $T$ for Boltzmann-derived Onsager coefficients of metals\cite{PhysRevB.64.224519}, one finds\cite{PhysRevB.70.054503}
\begin{equation}
\nu\approx - \frac{\pi^2}{3e} k_B^2 T \frac{\partial 
\mu_H
}{\partial\mu}
\label{eq:Mottnu}
\end{equation}
where $\mu_H=\LL_{11}^B / \LL_{11}$ is again the Hall mobility%
\footnote{This `Mott formula' for the Nernst coefficient is often written using the Hall angle $\tan (\theta_H) / B = \mu_H$ with the magnetic field $B$.}%
\footnote{Alternatively, the temperature-derivative of the Hall mobility can be seen as a source for a finite Nernst signal, $\propto T\partial\mu_H/\partial T$\cite{Sun_CoSb3}.}.
As for the Seebeck coefficient, we find the link
\eref{eq:Mottnu} between transport of charge ($\mu_H$) and entropy ($\nu$) to hold in the low-$T$ saturation regime (in this case $T<T_{R_H}^*$). In other words, the lowest order terms in the low temperature expansions, Eqs.~(\ref{eq:zero1}-\ref{eq:zero1B}),
fulfill \eref{eq:Mottnu}. In this sense,
the saturation of both $\sigma_{xx}$ and $\sigma_{xy}^B$ dictates the Nernst coefficient to vanish linearly for $T\to 0$.
This behavior---otherwise typical for metals\cite{0953-8984-21-11-113101}---is indeed experimentally observed in correlated narrow-gap semiconductors, see \fref{fig:materials} (h) for the example of FeAs$_2$.
In metals, however, the variation of the charge and Hall conductivities with the chemical potential is usually small.
Then, \eref{eq:Mottnu} means that also the Nernst coefficient will be very small---a statement referred to as Sondheimer cancellation\cite{PhysRevB.64.224519,Sondheimer,0953-8984-21-11-113101}. Instead, as we have seen here, 
a changing chemical potential can notably manipulate the
residual conductivities of an incoherent semiconductor.

\subsubsection{Thermal conductivity and Lorenz ratio}
\begin{figure*}[!h!t]
    \includegraphics[width=0.95\textwidth]{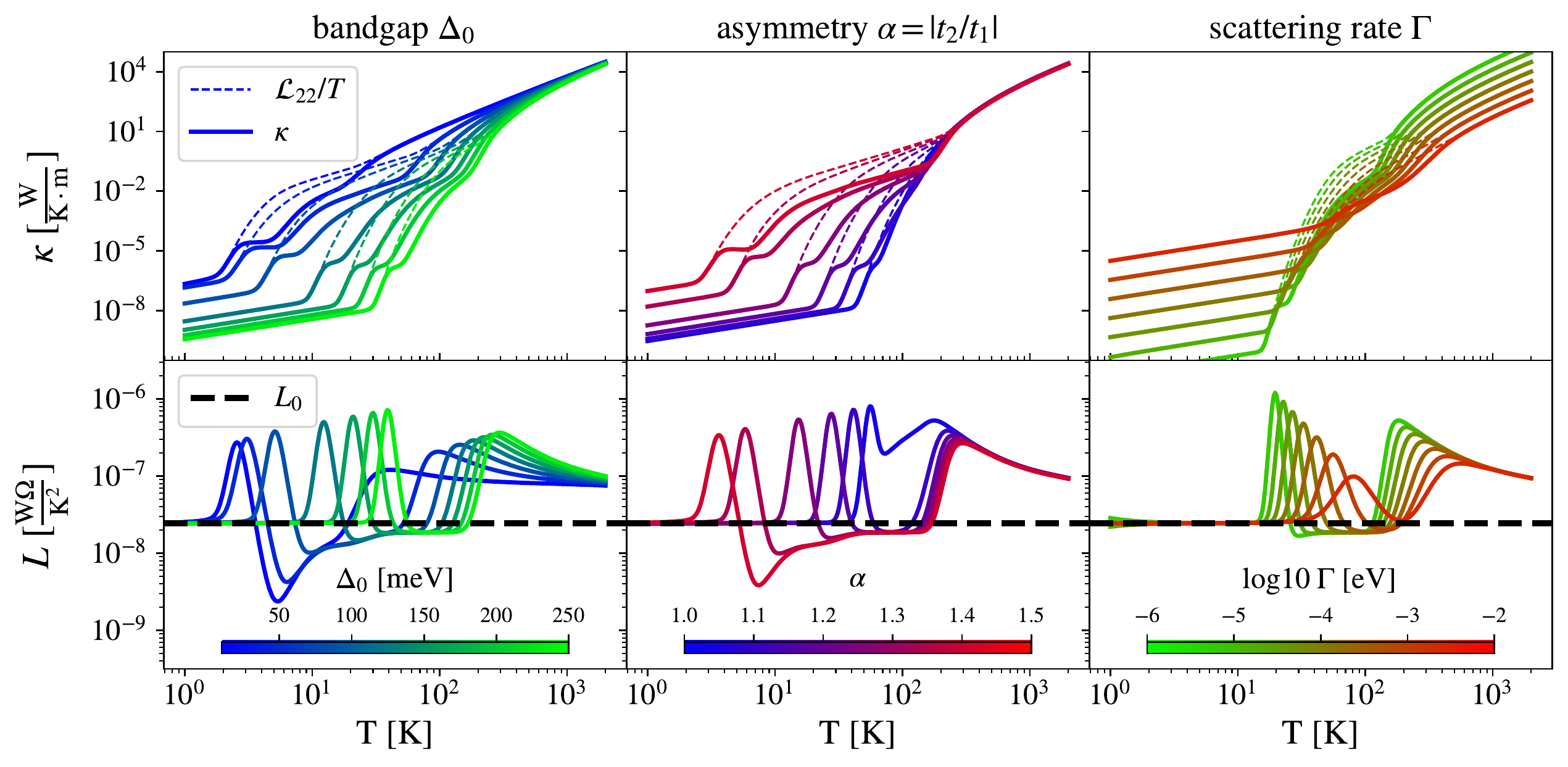}
    \caption{\textbf{Thermal conductivity and Wiedemann Franz law.} The same parameter sets as in \fref{fig:parameterscan} are used. In addition to the total thermal conductivity $\kappa$ (top row) we plot also the contribution $\mathcal{L}_{22}/T$  (dashed) individually. The Lorenz ratio (bottom row) converges to $L(T\rightarrow 0)  = L_0 =\frac{\pi^2 k_B^2}{3 e^2}$ (see text for details).
%    In fact, within our theory this relation must be fulfilled as the charge and heat current is carried by the same type of quasi particles. ... plus scattering is elastic
}
    \label{fig:wiedemannfranz}
\end{figure*}

Next, we discuss the electronic contribution to the thermal conductivity $\kappa$. In the low temperature $\rho$-saturation regime we find the linear behavior from \eref{eq:kappalowT}. Increasing the temperature results in various kinks and shoulders. 
Again, we can separate the influence of a changing chemical potential from the inner structure of the transport kernel functions:
While the pure $\LL_{22}$-contribution (first term in \eref{eq:ObservThermal}; dashed lines in \fref{fig:wiedemannfranz})
  only experiences the transition stemming from the chemical potential, the shoulders in the intermediate regime derive from the $\LL_{12}$-contribution (second term in \eref{eq:ObservThermal}). Combined to the Lorenz ratio $L$, see \eref{eq:Lorenz}, we find
a complex temperature dependence: In the zero temperature limit $L(T)$ clearly converges to the Sommerfeld value of the Lorenz number $L_0 = \frac{\pi^2 k_B^2}{3e^2}$.
This can also be confirmed via the low-$T$ expansion 
\begin{equation}
L= L_0 + \left( \frac{k_B}{e} \right)^2 \frac{16\pi^4}{45}\frac{5a^2-2\Gamma^2}{(a^2+\Gamma^2)^2} k_B^2 T^2 + \mathcal{O}(T^4),
\end{equation}
see \aref{sec:app_lowT} for more details.
This result is expected, since in our theory both the electrical and heat current are transported by the same carriers, subject to the same elastic scattering mechanism. If inelastic scattering effects, e.g., via an electron-phonon coupling, were to be included, 
this unison will be jeopardized\cite{PhysRevResearch.2.013148}.
Then,
the Wiedemann-Franz law can be strongly violated at low (but finite) temperatures, with $L$ being notably suppressed\cite{PhysRevB.99.085104}.

In the opposite limit, $T\to \infty$, 
we find $L(T\rightarrow\infty) = 0$. 
Note that this result assumes a temperature-independent scattering rate. For specific conditions on $\Gamma(T)$,
the Lorenz ratio converges to $L_0$ at high temperatures, as will be discussed in \sref{sec:tempgamma}.

\begin{figure*}[!h!t]
    \includegraphics[width=0.95\textwidth]{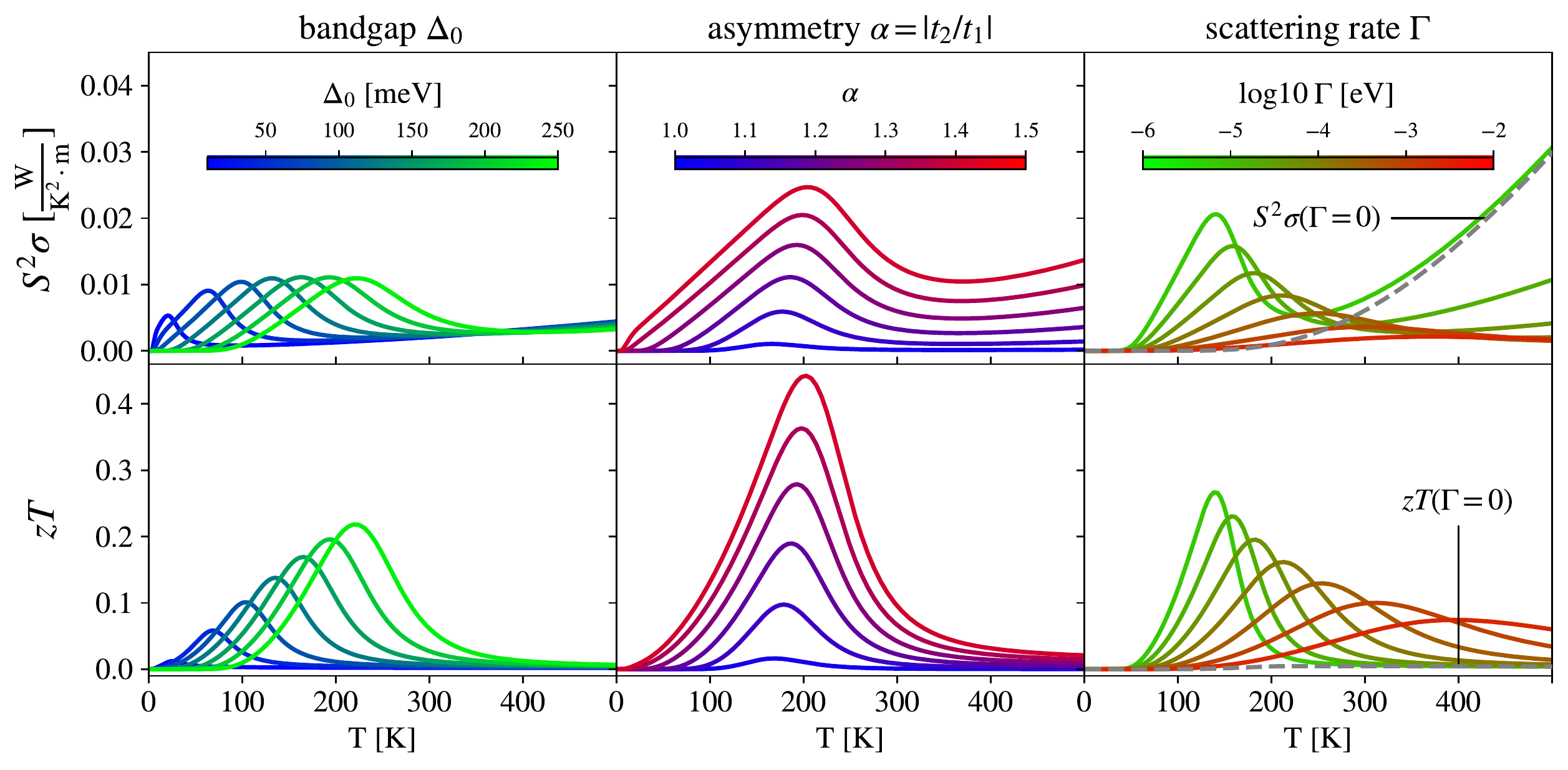}
    \caption{\textbf{Power factor and figure of merit.} The power factor $S^2\sigma$ (top row) is peaked at the transition of the first into the second activated regime of $\rho=1/\sigma$. For $T\to 0$, $S^2\sigma$ is suppressed due to the low temperature limit of $S(T)\sim T$.
    At intermediate temperatures, $S^2\sigma$ develops an important peak that is undetectable in Boltzmann theory (gray, dashed line).
    The high temperature increase in the powerfactor is nonphysical and disappears when a $T$-dependent scattering is included (see \sref{sec:fermiliquid}).
    The figure of merit $zT$ (bottom row) shows similar features but---due to the thermal-conductivity weighing---peak sizes are affected differently. Note: In $zT$ we included a constant phonon contribution of $\kappa_{\mathrm{ph}} = 10 \frac{\mathrm{W}}{\mathrm{K}^2\cdot\mathrm{m}}$. Again, the low temperature peak $zT$, indicative of potential merit in thermoelectric devices is completely missing in the Boltzmann regime (gray, dashed line).}
    \label{fig:powerfactor}
\end{figure*}

\subsubsection{Power factor and figure of merit}
\begin{figure}
    \centering
    \includegraphics[width=0.45\textwidth]{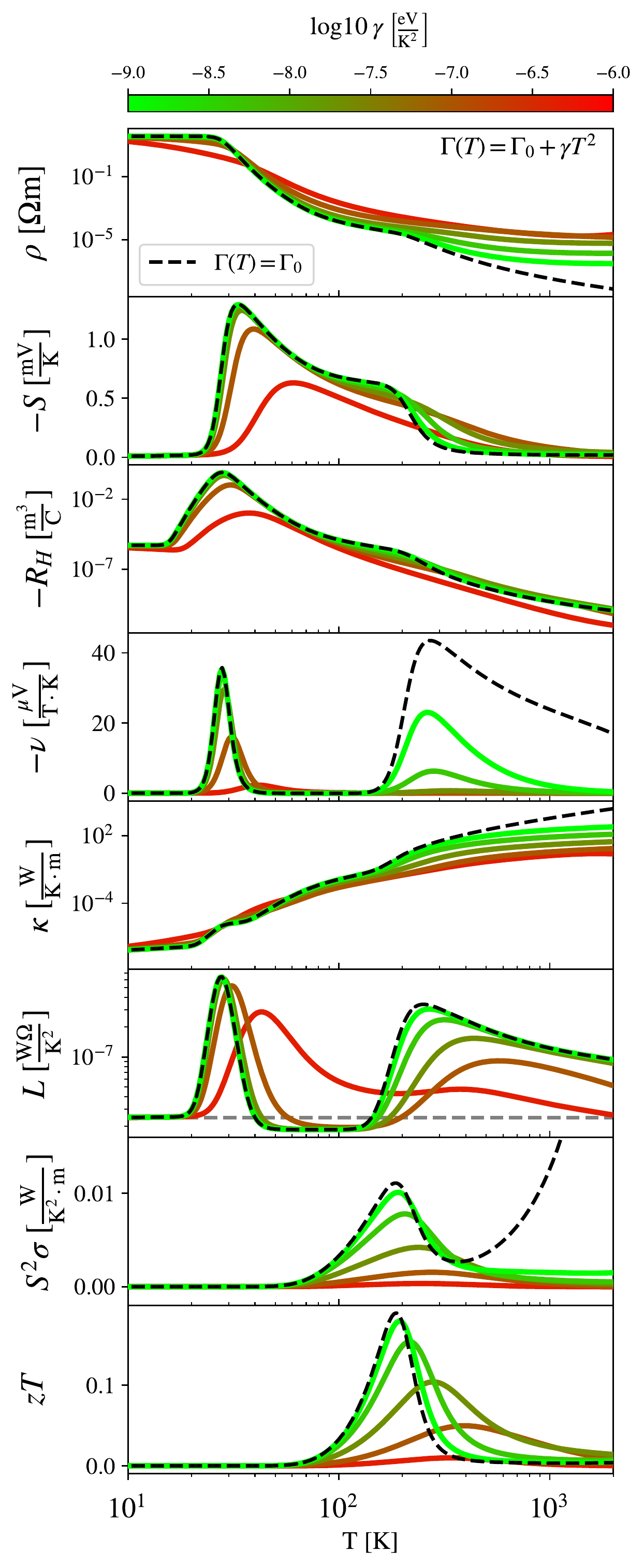}
    \caption{\textbf{Temperature dependent scattering rate.}
    Effects of $\Gamma(T) = \Gamma_0 + \gamma T^2$, with residual scattering $\Gamma_0=10^{-4}$eV and varying $\gamma$, for fixed bandgap $\Delta_0=200$meV and asymmetry $\alpha=1.2$. The dashed black lines are results for $\gamma=0$. The additional temperature dependence causes a resistivity upturn at high temperatures which also results in a smoothed Seebeck and Hall coefficient. This metallic trend directly removes the nonphysical upturn in the power factor while simultaneously causing the Lorenz factor $L$ to converge towards $L_0$ (horizontal, dashed, gray line).
    }
    \label{fig:Tsquared}
\end{figure}

Finally, we consider the thermoelectric power factor $S^2\sigma$ and the figure of merit $zT$, given by Eqs.~(\ref{eq:PF}-\ref{eq:FOM}) and shown in \fref{fig:powerfactor}.
Furthermore, in order to achieve somewhat realistic $zT$ values
we add a (dominant) phonon contribution to the thermal conductivity $\kappa = \kappa_\mathrm{e} + \kappa_{\mathrm{ph}}$ using an optimistic $\kappa_{\mathrm{ph}} = 10 \frac{\mathrm{W}}{\mathrm{K}^2\cdot\mathrm{m}}$.
First, we note that $S^2\sigma$ is seemingly large at the upper end of the shown temperature window.
However, this behavior again originates from the temperature-independence of the scattering rate---that we assume here for illustrative purposes. Indeed, $\Gamma=const.$ causes (for large $T$) a very small resistivity, see top row of \fref{fig:parameterscan}, that overcompensates the drop in the Seebeck coefficient.
In practice, the scattering rate itself is typically temperature dependent.  As explained below, see \sref{sec:fermiliquid}, a reasonable $\Gamma(T)$
 causes, both, the powerfactor and $zT$ to vanish quite rapidly at large temperatures, see \fref{fig:Tsquared} (lowest two panels).
We therefore focus on the lower temperature structure in $S^2\sigma$ and $zT$ in \fref{fig:powerfactor}, that is equally present when a realistic $\Gamma(T)$ is used.
The displayed peak in the powerfactor and $zT$
is the result of the usual compromise\cite{zlatic,behnia} between large $S$ and small $\rho$%
\footnote{Interestingly, this conventional trade-off was recently found to be broken in an ultra-thin oxide film near its Mott transition\cite{Katase_LaTiO3}.
}.
We find the optimal power factor to occur in the vicinity of $T_{\nu}^{\mu}$, the onset of the high-temperature crossover from the first into the second activated-$\rho$ regime. 
Peak temperatures move up (down) for a growing bandgap (scattering rate), while peak amplitudes benefit from larger gaps, larger asymmetry, but a smaller scattering rate.

As the ratio of powerfactor and thermal conductivity, $zT$ inherits its structure from the former, while the latter modulates the overall magnitude. Indeed, bandgap variations
keep the peak amplitude of $S^2\sigma$ essentially unchanged but move the peak position. The maximal $zT$ then increases for larger bandgaps, as the (here: electrical) thermal conductivity dwindles.
 Variations of the band asymmetry, instead, only change the size of the $zT$ maximum, while increasing scattering rates lower the peak amplitude and shift it to higher temperatures.
 
Importantly, the just described peak in, both, $S^2\sigma$ and $zT$ is {\it absent} when the Boltzmann approach is applied to the band structure (dashed grey lines in the right panels of \fref{fig:powerfactor}).
Indeed, we find that maximal thermoelectric performance is realized in the temperature range delimited by $T_{\Delta}^{\mu}$ and $T_{\nu}^{\mu}$. These characteristic scales are
driven (see above) by changes in the chemical potential $\mu$, caused by the finite lifetimes of conduction and valence states---an effect beyond mere thermal activation. As a consequence, assessing the potential of narrow-gap semiconductors for thermoelectric applications  on the basis of Boltzmann theory applied to coherent electronic band structures is virtually meaningless.
A (high-throughput) screening of materials\cite{C2CP41826F,PhysRevX.1.021012,C5TC04339E} that neglects finite electronic lifetimes of intrinsic carriers may miss potentially favorable compounds.

\subsection{Temperature dependent scattering rate}
\label{sec:tempgamma}
In the previous section, we held the scattering rate $\Gamma$ constant to unravel the prototypical variations of transport observables with respect to gap, particle-hole asymmetry and the scattering rate itself.
Setting  $\Gamma(T) = \Gamma_0$ led to some effects not observed in experiments. In particular, the resistivity became vanishingly small in the intermediate to high temperature regime ($\frac{\rho(T<T_\rho^*)}{\rho(T>T_{\nu}^{\mu})} \sim 10^5$). Indeed, only when temperature reaches a value greater than the system's band-width, the resistivity starts again to increase (not shown). 
Experiments probing narrow-gap semiconductors, however, witness
an insulator-to-metal crossover above a temperature that is still small with respect to the charge gap\cite{NGCS}.
In FeSi ($\Delta\sim 50$meV$=k_B\times 580$K), for example, the slope of the resistivity turns positive above 300K\cite{Bocelli_FeSi}, while in optical spectroscopy for FeSb$_2$ ($\Delta\sim 30$meV=$k_B\times 350$K)  a Drude-like peaks starts developing at around 100K\cite{perucci_optics,Homes2018}.
Clearly this metallization is beyond mere thermal activation of carriers across the charge gap. Theoretically, this phenomenon has been attributed to incoherent spectral weight spilling into the gap and was advocated to derive from electronic correlation effects\cite{jmt_fesi} or thermal disorder\cite{0034-4885-60-11-003,Delaire22032011}.
In the correlations' picture, the Hund's rule coupling drives a scattering rate that grows quadratically with temperature\cite{jmt_fesi,jmt_hvar}.
Therefore, we will restrict ourselves in the following to scattering rates with a polynomial temperature dependence.

\subsubsection{General considerations}
\label{sec:nonphysical}
Without a growing scattering rate the Lorenz ratio $L$ approaches zero in the high temperature limit and an unreasonably large power factor $S^2\sigma$ appears in the intermediate temperature regimes, see the high-$T$ upturn in \fref{fig:powerfactor}.
If, instead, we consider a residual scattering rate plus a term with a polynomial temperature dependence, $\Gamma(T) = \Gamma_0 + \gamma T^\eta$, where $\gamma>0$,  $\eta > 0$, the argument $z$ of the polygamma functions $\psi_i(z)$ becomes
\begin{equation}
    z(T) = \frac{1}{2}+\left[\frac{\Gamma_0 + ia}{2\pi k_B T} + \frac{\gamma T^\eta}{2\pi k_B T}\right].
    %\\ &= \frac{1}{2} + \lim_{T\rightarrow \infty} \frac{\gamma T^{\alpha-1}}{2\pi k_B}.
\label{eq:zt}
\end{equation}
Scattering rates that increase slower than linearly ($\eta<1$) lead to arguments that converge to $z(T\rightarrow\infty)=\frac{1}{2}$; exact linear behavior leads to  $z(T\rightarrow\infty)=\frac{1}{2} + \frac{\gamma}{2\pi k_B}$ while $\eta>1$ leads to a diverging $z(T)$.

In the first two cases the Lorenz ratio simplifies in leading order to $L(T) \sim \mathcal{O}( T^{2\eta-2})$. $\eta<1$ therefore implies a vanishing Lorenz ratio while $\eta=1$ implies some saturation value $L(\gamma)$, which, numerically, is generally orders of magnitude smaller than $L_0$.
If the scattering rate increases faster ($\eta>1$), the same Taylor series of the polygamma functions that was applied in the zero temperature limit (see Appendix~\ref{sec:app_lowT}) can be employed. Consequently, the high temperature limit is identical to the low temperature limit and thus $L(T\rightarrow\infty) = L_0$.

\subsubsection{Fermi liquid-like scattering}
\label{sec:fermiliquid}
Dynamical mean-field theory calculations suggest that the scattering rate evolves quadratically with temperature
for, both, Kondo insulators\cite{LRT_Tstar} and $d$-electron-based narrow-gap semiconductors\cite{jmt_fesi,jmt_hvar}.
From here on, we therefore assume a Fermi-liquid-like
\begin{equation}
    \Gamma(T) = \Gamma_0 + \gamma T^2.
    \label{eq:scatterT2}
\end{equation}
\fref{fig:Tsquared} illustrates how the additional scattering term changes the transport for a range of $\gamma$-values (shades green to red)
compared to $\gamma=0$ (dashed black line), for a fixed bandgap $\Delta_0=200$meV, residual scattering $\Gamma_0=10^{-4}$eV, and asymmetry $\alpha=1.2$.

Overall, the increased scattering smoothes all considered quantities. A minimal $\gamma$ is sufficient to suppress the high temperature shoulder in $\rho$ at $T_{\nu}^{\mu}$ and causes a slight upturn at high temperatures. The saturation regime is instead stable up to rather large $\gamma$.
Naturally, the added scattering term only notably influences transport above temperatures for which $\gamma T^2\sim\Gamma_0$.
In this vein the high-temperature shoulder of the Seebeck coefficient is smoothed out and parts of the signal is pushed up in temperature. Quite generally, the increased scattering leads to less pronounced peaks which are shifted to higher temperatures. Since the shoulder in $S$ at $T_{\Delta}^{\mu}$ was responsible for the strong signal in the power factor and the figure of merit (see above), significant qualitative changes are expected for $\gamma>0$:
Besides the suppression of the nonphysical increase of the power factor at high temperatures, both $S^2\sigma$ and $zT$ are equally attenuated for $\gamma>0$ and their peaks shift up in temperature, as expected.
Again, we included in $zT$ a phonon contribution, $\kappa_{\mathrm{ph}} = 10 \frac{\mathrm{W}}{\mathrm{K}^2\cdot\mathrm{m}}$, to the thermal conductivity.
As already seen in the $\Gamma_0$-scan in \fref{fig:parameterscan}, a noticeable 
suppression is observed for the high-temperature peak of the Nernst coefficient. 
The second, low-$T$ peak in the Nernst coefficient (that is absent in Boltzmann approaches) is instead relatively stable with $\gamma$ as it occurs at low enough temperatures $T\sim T_\rho^*$.
As discussed in \sref{sec:nonphysical} the Fermi-liquid like scattering rate enforces that the high-temperature limit of the Lorenz ratio converges to $L_0$. This is evident in \fref{fig:Tsquared}: For the largest scattering rate, $L$ reaches $L_0$ within the shown temperature window.

\section{Modeling Materials}\label{sec:results2}
Having established an understanding of how relevant parameters drive 
changes in transport observables, we now turn 
to material specific simulations.
While still focusing on a minimal description, we
attempt to reproduce the temperature profiles of transport properties in selected narrow-gap semiconductors, as well as trends among them.

\subsection{FeSb$_2$: Characteristic temperature scales without impurity states}
\label{sec:fesb2intrinsic}
First, we discuss the result for FeSb$_2$ shown in the front \fref{fig:prototypical}. FeSb$_2$ is a correlated semiconductor\cite{0295-5075-80-1-17008,PhysRevB.72.045103,perucci_optics,PhysRevB.88.245203} with a narrow gap of $\Delta \gtrsim 30$meV, as extracted from activation-law fits of the resistivity\cite{0295-5075-80-1-17008,PhysRevB.88.245203} or the magnetic susceptibility\cite{PhysRevB.74.195130,Takuya-Deguchi2020MT-MN2019015}.
Consistent with $GW$ and $GW$+DMFT simulations\cite{jmt_fesb2,PhysRevResearch.2.023190}, we model FeSb$_2$ with a
non-interacting gap $\Delta_0=60$meV and an effective mass enhancement $Z^{-1}=2$.
We find that a small asymmetry $\alpha = 1.06$ mimics the material well. 
Finally, we assume a scattering rate of the form \eref{eq:scatterT2}, where the parameters of the residual scattering $\Gamma_0$ and the prefactor $\gamma$ of the quadratic term are adjusted by hand.
We find that best agreement with experiment is reached for $\Gamma(T) = 5\cdot10^{-5}\mathrm{eV}+ 10^{-7} \frac{\mathrm{eV}}{\mathrm{K}^2}T^2$, see \fref{fig:prototypical}. This scattering rate is quite realistic: Dynamical mean-field calculations for the related correlated narrow-gap semiconductor FeSi\cite{jmt_fesi,jmt_hvar} yield comparable values.
Having optimized the electronic structure parameters so that the simulated resistivity qualitatively follows the experiment, the temperature profiles of all other transport observables for FeSb$_2$ automatically fall into place, see \fref{fig:prototypical}. 
The approach therewith in particular verifies the experimentally observed correlation between features in different response functions:

At low temperatures, the onset of resistivity saturation at $T_\rho^*$ is accompanied by a peak in, both, the Hall and the Nernst coefficient at the identical temperature and a peak in the Seebeck coefficient at a slightly higher temperature---all of which is congruent with experiment. 
This low-$T$ behavior of transport properties is encoded in the linear response kernel functions. Agreement with experiment confirms that our approximations for the kernels---linearized self-energy, omission of vertex corrections---conserves the essential physics.
Instead, in previous modelings of FeSb$_2$, based on semi-classical approaches\cite{FeSb2_Marco,doi:10.7566/JPSJ.88.074601,PhysRevB.103.L041202}, resistivities and the Hall coefficient either diverged at low $T$ or had to be suppressed by impurity states, e.g., by forcing the chemical potential into the conduction band.
An alternative scenario for residual conduction in FeSb$_2$ could be provided by
the recent observation of metallic surface states\cite{PhysRevResearch.2.023190,Xu15409}. Whether these weakly dispersive states can account 
for the typical low-$T$ characteristics across all transport observables remains to be seen. We also note that for the topological insulator SmB$_6$ conduction by surface states and residual bulk conduction from finite lifetimes coexist\cite{LRT_Tstar}.

Moving to higher temperatures, 
the resistivity exhibits two distinct activation regimes. We
find the shoulder
in between, $T_{\Delta}^{\mu} < T < T_{\nu}^{\mu}$, to be driven by changes in the chemical potential. 
Therefore, if the chemical potential only accounts for the thermal broadening of excitations---as commonly done in the context of Boltzmann approaches for band theory methods---all structure at intermediate temperatures is lost (gray, dotted lines in \fref{fig:prototypical}).
If instead, the chemical potential, e.g., via \eref{eq:digammocc}, accounts for the scattering rate, Boltzmann simulations do capture 
the intermediate-$T$ features
(see black, dashed lines in \fref{fig:prototypical}). 
Alternatively, the temperature dependence of the chemical potential can be engineered by assuming in-gap impurity states\cite{FeSb2_Marco,doi:10.7566/JPSJ.88.074601,PhysRevB.103.L041202}.
Given that transport observables exhibit three to four distinct regimes, phenomenological modellings actually used up to three impurity levels to properly guide the chemical potential\cite{FeSb2_Marco}. 
In our description, no impurity states are required: According to the 
presented phenomenology for transport in narrow-gap semiconductors,
the intriguing temperature dependence in FeSb$_2$ exclusively originates from its {\it intrinsic} electronic structure.
Instead of being set by {\it explicit} energy levels inside the gap, characteristic temperatures emerge through the interplay of relevant energy scales: the gap, the hopping, temperature, and---crucially---the scattering rate.

In the following section, we investigate the influence of potential impurity states in more detail for FeSb$_2$. Thereafter, we will see that {\it explicit} impurity states are not fully out of the picture for other materials, but likely account for at least some aspects of conduction at intermediate temperatures in semiconductors with gaps $\Delta>50$meV.

Before, however, a few comments regarding thermoelectricity in FeSb$_2$ are in order. While our approach neatly captures the {\it temperature profile} of transport observables, we do not reproduce the large {\it amplitude}
of the Seebeck and Nernst coefficient. 
In fact, this is expected, as FeSb$_2$ violates the upper bound, $|S|\leq \Delta/T$, for a thermopower exclusively driven by electron diffusion\cite{jmt_fesb2}.
As alluded to in the introduction, this riddle was successfully solved\cite{jmt_fesb2,MRC:8871060,Takahashi2016,FeSb2_Marco,doi:10.7566/JPSJ.88.074601,PhysRevB.103.L041202} by attributing the colossal amplitude to the phonon-drag effect. Simply speaking, 
the thermal gradient also leads to a non-equilibrium phonon distribution. Working to equilibrate the thermal gradient, phonons then scatter with electrons dominantly towards the cold end of the sample, which is also the direction of the net electronic diffusion. Thereby momentum is constructively injected
into the electronic subsystem, significantly boosting thermoelectric effects. This well-known phenomenon\cite{PhysRev.96.1163}
continues to receive renewed interest, in the context of electronic correlations\cite{FeSb2_Marco,PhysRevB.103.L041202,NGCS} (the effect is large when coupling to heavy electrons), the phonon-engineering pathway to efficient thermoelectrics\cite{Zhou01122015}, or both\cite{Katase_LaNiO3_hetero}.
Crucially for our argument here, this phonon-enhancement of the electric response is expected to be smooth in temperature, so as to not produce additional features in transport observables. 
Indeed, while experimental peak-amplitudes cover almost an order of magnitude, $S^{\max}\sim 5 - 45 \frac{\mathrm{mV}}{\mathrm{K}}$ across different samples\cite{0295-5075-80-1-17008,sun_dalton,PhysRevB.88.245203,Du2021}, 
the corresponding characteristic temperature profiles are almost identical.
Crucially, Pokharel \etal~\cite{MRC:8871060} and
Takahashi \etal~\cite{Takahashi2016}
demonstrated that the phonon-drag in FeSb$_2$ can be consistently suppressed by geometric constraints.
With their severely limited phonon mean-free path,
polycrystalline samples are then expected to yield Seebeck amplitudes compatible with the purely electronic diffusion
simulated here. Indeed, experimental peak-amplitudes
for polycristalline samples, $S^{\max}\sim\mathcal{O}(0.1-1)\frac{\mathrm{mV}}{\mathrm{K}}$\cite{bentien:205105,MRC:8871060,Sanchela2015205,Takahashi2016} are comparable to our modelling, see \fref{fig:prototypical}.
With the phonon-drag thus mainly scaling the amplitude of the thermoelectric response, previous modellings including this effect had to explicitly introduce in-gap impurity levels\cite{FeSb2_Marco,doi:10.7566/JPSJ.88.074601,PhysRevB.103.L041202}, to generate the experimentally evidenced characteristic temperature scales.
Here, we showed that the electron diffusion part of the Seebeck and Nernst coefficient has the correct temperature profile without the need for {\it ad hoc} in-gap levels---provided that finite lifetimes of intrinsic carriers are accounted for consistently.

\begin{figure}[!ht]
    \centering
    \includegraphics[width=0.45\textwidth]{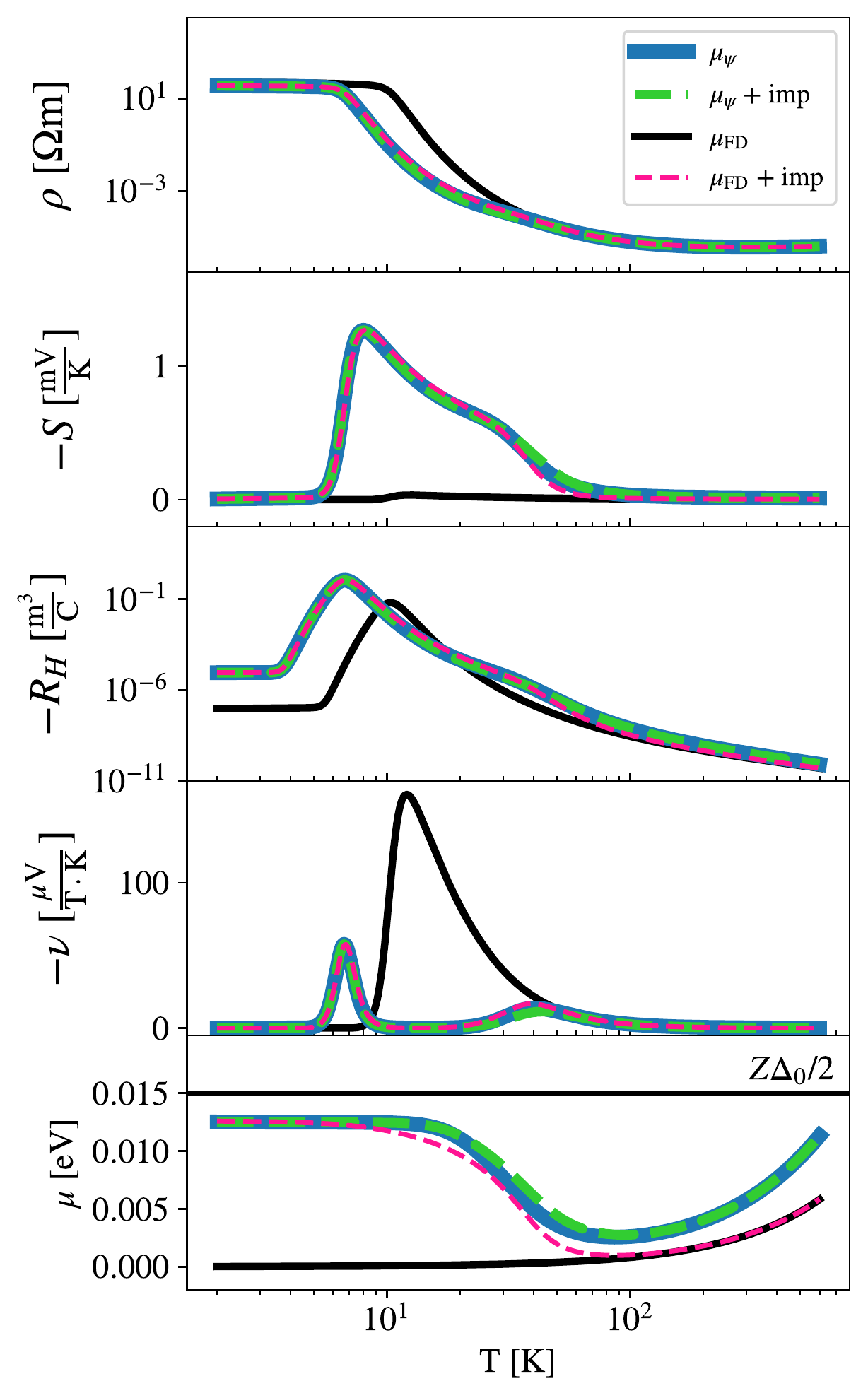}
    \caption{\textbf{FeSb$_2$ and impurity in-gap states.} We imitate the chemical potential determined via \eref{eq:digammocc}---$\mu_\psi$---with the Fermi-Dirac distribution through the presence of an explicit impurity level---$\mu_{\mathrm{FD}} + \mathrm{imp}$---in the vicinity of the conduction band. We employ the same parameters as in \fref{fig:prototypical}. Additionally, an impurity donor level with density $\rho_D= 5\cdot 10^{-6} \frac{1}{\mathrm{unit\;cell}}$, degeneracy $g=1$ is positioned at $E_D = 26$ meV below the conduction band (before renormalization via $Z$). The chemical potential is then adjusted according to
    $N \equiv n_{occ}(\mu) - \frac{\rho_D}{1 + g \cdot f(\mu - E_D)}$. For comparison we also show the (very different) chemical potential determined by the Fermi-Dirac distribution without the donor level---$\mu_{\mathrm{FD}}$---and the 
    digamma-computed chemical potential with the additional impurity---$\mu_\psi + \mathrm{imp}$---which is indistinguishable from $\mu_\psi$. All shown transport quantities employ the full {\texttt{\textsc{LinReTraCe}}} kernels, Eqs.~\ref{LRT_k11}-\ref{LRT_k22B}.} 
    \label{fig:prototypical_mimic}
\end{figure}

\subsection{FeSb$_2$: Explicit impurity states?}
To strengthen the argument that in-gap impurity states are not crucial for an understanding of transport properties of FeSb$_2$, we study the explicit inclusion of such states.
In \fref{fig:prototypical_mimic} (bottom panel) we compare several ways to obtain the needed chemical potential: (i) $\mu_\psi$ (solid blue) indicates an occupation determined via \eref{eq:digammocc}, that accounts for, both, thermal broadening and the finite lifetimes of valence and conduction states (data reproduced from \fref{fig:prototypical}); (ii) $\mu_{FD}$ (black) that only includes thermal broadening via the Fermi-Dirac distribution; (iii) $\mu_{FD}+\hbox{imp}$ (pink dashed) in which an in-gap impurity level has been designed to mimic $\mu_\psi$; and (iv) $\mu_\psi+\hbox{imp}$ (green dashed) in which the same impurity level is added in the presence of finite lifetimes of intrinsic states.

Clearly, the Fermi-Dirac description of the chemical potential (black), in which the chemical potential converges towards the mid-gap point (origin of energy) for $T\to 0$, yields very different transport functions (upper panels). Their temperature profiles do not agree with experiment, cf.\ \fref{fig:prototypical} (insets). Still, the resistivity and the Hall coefficient saturate (albeit at values different than in the "$\mu_{\psi}$" scenario) and the Nernst and Seebeck coefficient vanish for $T\to 0$, since these low temperature properties stem from the employed {\texttt{\textsc{LinReTraCe}}} kernels, Eqs.~\ref{LRT_k11}-\ref{LRT_k22B}.
As a consequence, if the chemical potential $\mu_\psi$---that drives both, the features at intermediate temperatures and influences the saturation values---could be mimicked by other means, transport properties will be very similar to the intrinsic "$\mu_{\psi}$" picture.
And, indeed, imitating the temperature dependence of $\mu_{\psi}$ through the inclusion of a single explicit donor level (at an energy $E_D=26$meV, degeneracy $g=1$ and density $\rho_D=5\cdot10^{-6}$ per unit-cell),
these "$\mu_{FD}+\hbox{imp}$"-results (pink dashed) are  very close to the
$\mu_\psi$ results.
In turn, if we include the same impurity level on top of the lifetime effects that drive $\mu_\psi$ (a combination labelled "$\mu_\psi+\hbox{imp}$" (green dashed) in \fref{fig:prototypical_mimic}), nothing much happens. In other words:
if finite lifetimes of intrinsic valence and conduction states are properly accounted for, extrinsic in-gap states have little on transport properties in FeSb$_2$.
This finding strengthens our alternative scenario in which the driver of the characteristic temperature profile in transport properties is the scattering rate.

\begin{figure*}
    \centering
    \includegraphics[width=0.95\textwidth]{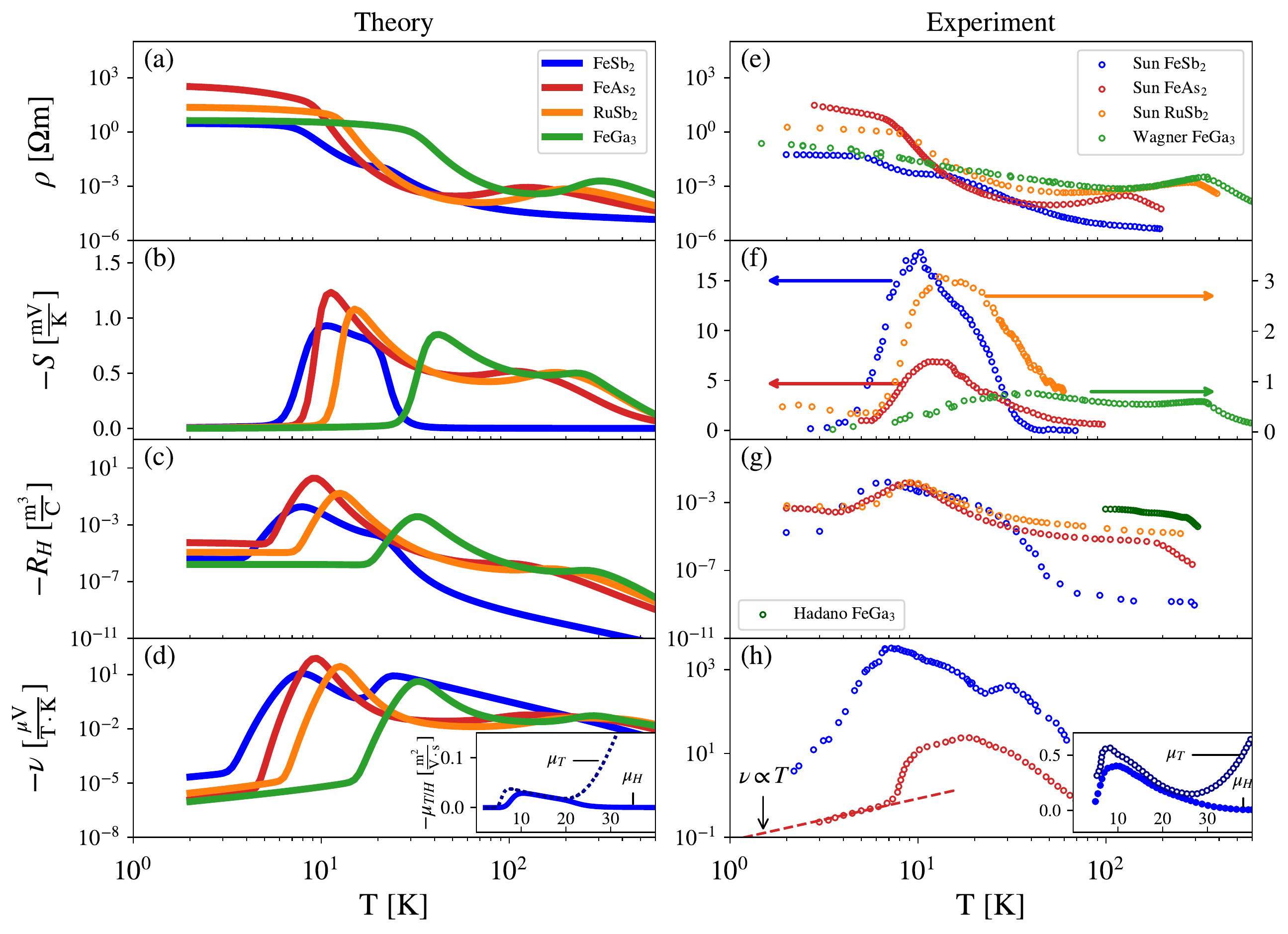}
    \caption{\textbf{Modelling of intermetallic narrow-gap semiconductors.} The experimental temperature profile of transport observables of
    FeSb$_2$\cite{PhysRevB.88.245203}, 
    FeAs$_2$\cite{PhysRevB.88.245203}, 
    RuSb$_2$\cite{sun_dalton} and
    FeGa$_3$ ([100] orientation) \cite{PhysRevB.90.195206,JPSJ.78.013702} is simulated with finite lifetimes and a single donor level at a distance $E_D$ below the conduction band.
    Lifetimes dominantly determine transport at intermediate to low temperatures. Instead, the donor level virtually only affects higher temperatures near $T_{\Delta}^{\mu}$ and $T_{\nu}^{\mu}$. In particular, in congruence with experiment, the level accounts for the metallic slope in the resistivity seen in all materials other than FeSb$_2$.
    To limit the number of adjustable parameters, bandgaps were fixed to experimental values: using $Z=0.5$ for $\Delta_{\mathrm{FeSb}_2} = Z\times\Delta_{0,\mathrm{FeSb}_2} = 30$meV and $Z=1$ in $\Delta_{\mathrm{FeAs}_2} = 200$meV; $\Delta_{\mathrm{RuSb}_2} = 290$meV; $\Delta_{\mathrm{FeGa}_3} = 500$meV. All modelling parameters are listed in \tref{tab2}. Jointly, finite lifetimes and the impurity level yield an accurate description without the need for other electronic structure details.
    }
    \label{fig:materials}
\end{figure*}

\subsection{Related materials: FeAs$_2$, FeGa$_3$, RuSb$_2$}
\label{sec:materials}

We now extend our transport study to other materials. In the right column of \fref{fig:materials} we reproduce experimental data of various intermetallic semiconductors. In order of increasing gap: FeSb$_2$\cite{PhysRevB.88.245203}, FeAs$_2$\cite{PhysRevB.88.245203}, RuSb$_2$\cite{sun_dalton} and FeGa$_3$\cite{PhysRevB.90.195206,JPSJ.78.013702}. For all considered compounds, the charge gap can be extracted directly from the high-temperature behavior of the resistivity (or optical data). With the exception of FeSb$_2$ ($Z^{-1}=2$), we do not apply a quasi-particle renormalization ($Z=1$). Indeed,
larger hybridization-gap semiconductors are expected to exhibit
less correlation signatures\cite{NGCS}
and also the substitution of a $3d$ transition-metal with its $4d$ homolog will reduce correlation effects\cite{jmt_hvar},
as explicitly shown for Fe$_{1-x}$Ru$_x$Sb$_2$\cite{cava2013fesb2,Du2021FeSb2}.

While the resistivity of FeSb$_2$ only displays a shoulder at $T\sim T_{\Delta}^{\mu}$ (successfully modelled with $\Gamma(T)$), a distinct peak can be observed in the three other materials.
The metallic slope, $\partial\rho/\partial T>0$, at temperatures below said peak \emph{cannot} be replicated with a chemical potential that is driven by finite lifetimes through Eq.~\ref{eq:digammocc} alone. Indeed, the transition of the chemical potential must occur more abruptly in temperature, making explicit impurity states a necessity to achieve agreement with experiment.
We find that deploying a single donor level near the conduction band allows us to reproduce the qualitative behavior of all considered materials across all considered transport observables. 

\paragraph{Resistivity.} For FeSb$_2$ this leads to minor improvements in $T_{\Delta}^{\mu}$ and $T_{\nu}^{\mu}$ of the initial fit (\fref{fig:prototypical}).
For Fe/RuAs$_2$ and FeGa$_3$ the 
engineered chemical potential causes the resistivity to decrease when cooling below their $T_{\nu}^{\mu}$ (metallic slope), before it rises again to enter the second activated region and, eventually, the saturation regime below $T_\rho^*$. The overall agreement is astounding: with a single impurity level characteristic temperatures, qualitative features and even relative amplitudes between various materials can be accurately modelled.%
\footnote{The only noticeable deviation constitutes FeGa$_3$ where the chosen experiment exhibits a less clear transition between an activated and a saturated region. See, however, the c-axis resistivity in  Ref.~\cite{JPSJ.78.013702}.}
\paragraph{Seebeck coefficient.}
Unsurprising from the previous analysis and the optimal parameters listed in \tref{tab2} the magnitude of the theoretical Seebeck coefficients (Fig.~\ref{fig:materials}b) do not differ significantly in the modelling. For all materials considered we observe a peak amplitude in the range of $\left\vert S^{\max} \right\vert = 0.8 - 1.3 \frac{\mathrm{mV}}{\mathrm{K}}$ positioned at a respective $T_S^{\max}$ slightly above the saturation temperatures $T_\rho^*$ of Fig.~\ref{fig:materials}a. 
While lacking the phonon-drag boost to the Seebeck magnitude, our treatment
still captures quite well, both, the dominant peak's position at $T_S^{\max}$ and the high temperature shoulder (peak) at $T_{\Delta}^{\mu}$ for FeSb$_2$ (FeGa$_3$).
Unfortunately, no data for FeAs$_2$ and RuSb$_2$ is available for higher temperatures: For them, we anticipate an additional Seebeck feature where the respective resistivities are peaked.
\paragraph{Hall coefficient.}
The agreement to experiment for the Hall coefficient, Fig.~\ref{fig:materials}g, is comparable to that of the resistivity: As in the experiments, the position of the peak in $R_H$  virtually coincides with the $T_\rho^*$ crossover in the resistivity. Equally the hierarchy across materials is captured for the saturation value, $R_H(T\to 0)$, and also the decay at higher temperatures mirrors the experiment. Clearly, the high-temperature shoulder is connected to the resistivity peak at $T_{\Delta}^{\mu}$. We therefore expect $R_H$ of RuSb$_2$ to similarly drop if temperatures slightly beyond the shown experimental range were probed.
\paragraph{Nernst coefficient.}
For the Nernst coefficient less experimental data is available, see Fig.~\ref{fig:materials}h. For the cases of FeSb$_2$ and FeAs$_2$ the qualitative agreement between simulation and measurements is reasonably good. Of course, what has been said about the Seebeck coefficient of FeSb$_2$ also applies to its Nernst signal: It is substantially boosted by the phonon drag\cite{FeSb2_Marco,PhysRevB.103.L041202}, limiting us to discussing the qualitative temperature profile%
\footnote{Contrary to $S$, $\nu$ depends on the lattice constant and scales according to $\nu\propto a_\mathrm{lattice}$, see Appendix~\ref{sec:app_latticescaling}. Using instead of our generic $a=1$\AA, a realistic lattice constant, FeSb$_2$'s Nernst amplitude is, in principle, not out of reach of the electron diffusion picture.
}.
Not suffering from this intricacy, clearer agreement is seen for FeAs$_2$: There, the experimental low temperature signal neatly follows the linear behavior $\nu\propto T$ (dashed line) as derived in \eref{eq:zero2} and illustrated in Fig.~\ref{fig:materials}d.

For FeSb$_2$, we also indicate the Hall and thermal mobilities of \eref{eq:numobilities} in the insets of Fig.~\ref{fig:materials}d/h. The experimental data\cite{PhysRevB.88.245203} qualitatively matches the theoretical prediction: At high temperatures a divergence between $\mu_T$ and $\mu_H$ is observed, giving rise to FeSb$_2$'s smaller Nernst peak at $T\approx 40$K. Below, at intermediate temperatures, $T\approx 20$K, the two mobilities almost coincide (i.e., $\nu$ is suppressed). At low temperatures, $T\approx 10$K, again a slight mismatch occurs, giving rise to the prominent low temperature peak. 

\subsection{Perspective}
The previous section made clear that
with reasonable scattering rates and (for larger gap systems) an explicit impurity in-gap level,
  all experimental transport coefficients can be qualitatively matched with an essentially featureless band structure. This emphasizes the notion that most---if not all---of the relevant transport physics in narrow-gap semiconductors is purely determined by the interplay of the gap, the chemical potential profile  (shaped by temperature, carrier lifetimes and, potentially, impurity states) and
  the scattering rate. Electronic structure intricacies, such as details of the band structure beyond the gap value and optical transition elements, 
  all seem to play only a secondary role.
Further, we evidenced that, at low temperatures ($T_{R_H}^*<T_{\rho}^*<T_S^{\max}$), features  are controlled by the scattering rate through the (quantum = beyond-Boltzmann) transport kernels, whereas the higher-temperature features ($T_{\Delta}^{\mu}<T_{\nu}^{\mu}$) 
are determined through the behavior of the chemical potential,
which can be driven by the scattering rate of intrinsic carriers as well as by extrinsic impurity in-gap states.

Future extensions of the presented formalism (and software package\cite{LRT}) could include the ability to describe phonon-drag contributions to thermoelectric observables.
This advance could remedy our current inability to quantitatively match the amplitude of the Seebeck coefficient, in particular of single crystalline FeSb$_2$. 
Furthermore, an inclusion of (topological) surface conduction and in-depth comparisons of their importance vis-à-vis the discussed bulk conduction is desirable, also in view of FeSb$_2$\cite{PhysRevResearch.2.023190,Xu15409}.
Finally, also anomalous bulk contributions could be included, following, e.g., the recent Ref.~\cite{PhysRevB.102.165151}.

\begin{table}[!h]
\begin{tabular}{l|ccccc|cc}
 & $Z$ & $\Delta_0$ & $\alpha$ & $\Gamma_0$  & $\gamma$   & $E_D$ & $\rho_D$ \\ 
 & & {[}meV{]} &  & {[}eV{]} & {[}$\frac{eV}{\mathrm{K}^2}${]}  & [meV] & [$\frac{1}{\mathrm{unit\;cell}}$]\\ 
 \hline
FeSb$_2$  & 0.5 & 60       &    1.02        & $1.5 \cdot 10^{-4}$             & $8 \cdot 10^{-7}$     & 20 &  $6\cdot 10^{-8}$    \\
FeAs$_2$  & 1   & 200  & 1.12                & $1.5 \cdot 10^{-5}$          & $3 \cdot 10^{-7}$        & 15  & $1.5\cdot 10^{-4}$                \\
RuSb$_2$  & 1   & 290   & 1.12                & $7 \cdot 10^{-5}$               & $2 \cdot 10^{-7}$     & 18 & $3.5 \cdot 10^{-4}$                     \\
FeGa$_3$  & 1   & 500   & 1.04               & $3 \cdot 10^{-4}$              & $2 \cdot 10^{-7}$ & 40 & $1.5 \cdot 10^{-3}$
\end{tabular}
\caption{{\bf Electronic structure parameters for  simulated materials}. Quasi-particle renormalization $Z$, bandgap $\Delta_0$, band asymmetry $\alpha=\left\vert t_2 / t_1\right\vert$, scattering rate coefficients in $\Gamma(T) = \Gamma_0 + \gamma T^2$ and a single donor level of concentration $\rho_D$ positioned at a distance $E_D$ below the conduction band. Gap sizes, $\Delta=Z\Delta_0$, are kept fixed to experimental values during the (manual) optimization procedure.}
\label{tab2}
\end{table}

\section{Summary and Conclusions}\label{sec:summary}

We conceptualized an efficient linear response transport formalism: A low-energy expansion of quasi-particle renormalizations
enabled performing frequency integrations in dominant Feynman diagrams {\it analytically}. This algorithmic innovation, implemented in the {\texttt{\textsc{LinReTraCe}}}\cite{LRT} package, 
allows 
accurate simulations down to temperatures where full Kubo calculations are cumbersome and Boltzmann techniques insufficient.
We applied the methodology to narrow-gap semiconductors and established a comprehensive phenomenology of their transport properties:

First, we analytically extracted low-temperature characteristics of various
transport observables:
In congruence with experiments, both, the resistivity and the Hall coefficient saturate at finite values for $T\to 0$.
The Seebeck and Nernst coefficients, instead, vanish linearly in the zero temperature limit, consistent with thermodynamic expectations. These behaviours are beyond the reach of semi-classical approaches like Boltzmann theory in the relaxation time approximation, highlighting the importance of a fully quantum mechanical description.

Next, we simulated transport properties as a function of temperature
for varying bandgap, particle-hole asymmetry and scattering rate. 
This survey establishes the prototypical temperature dependence of transport in narrow-gap semiconductors
to be structured by five emergent characteristic temperatures:
$T_{R_H}^* < T_{\rho}^* < T_S^{\max} < T_{\Delta}^{\mu} < T_{\nu}^{\max}$:
At high temperatures, $T>T_{\nu}^{\mu}$, the shape of all observables is controlled by the bandgap $\Delta_0$ and our equations yield results identical to Boltzmann approaches.
Upon cooling, $T_{\Delta}^{\mu}<T< T_{\nu}^{\mu}$, the system experiences a crossover from
the activated, Boltzmann-like regime 
to a renormalized activation region, $T_{\rho}^{*}<T< T_{\Delta}^{\mu}$, with an associated energy $\Delta_1<\Delta_0$.
We find this crossover to be driven by the chemical potential:
Finite lifetimes of valence and conduction states cause incoherent spectral weight to spill into the gap. 
Below a characteristic temperature, these incoherent carriers can no longer be neglected with respect to the charges that are thermally activated across the gap $\Delta_0$. In particle-hole asymmetrical systems, the chemical potential then has to adapt to preserve charge neutrality by moving to a position separated by only $\Delta_1$ from the top (or bottom) of the  valence (or conduction) band.
This {\it intrinsic} mechanism provides an explanation  alternative to the common {\it extrinsic} scenario in which the moving of the chemical potential (and the associated shoulder in the resistivity) is attributed to the presence of impurity in-gap states.
Finally, at low enough temperature, $T<T_{\rho}^*$ ($T<T_{R_H}^*$) the system enters the aforementioned lifetime-dominated regimes in which the resistivity (the Hall coefficient) saturates and thermoelectric observables vanish linearly.

In semi-classical approaches, the lifetime of excitations appears as a mere prefactor of, say, the conductivity. 
In the quantum formulation derived here, instead,
the scattering rate is a relevant control parameter that can compete with other energy scales of the problem.
Indeed, the emergence of all characteristic transport features is a direct consequence of the interplay of the charge gap and the scattering rate.
In other words, in our transport equations, thermal and lifetime broadening are described on an equal footing.

Crucial for potential applications, we find the {\texttt{\textsc{LinReTraCe}}} methodology to be essential to describe the temperature region where the thermoelectric powerfactor and the figure of merit is optimal. Materials discovery efforts based on conventional Boltzmann approaches, instead, are likely to overlook promising compounds.
To mend this shortcoming, we note that our methodology allows for an easy integration into already established code bases, heralding future
high-throughput material scans using, e.g., phenomenological scattering rates.

The established transport phenomenology further allowed us to fit experimental measurements and reverse engineer scattering rates---a task previously highlighted for Kondo insulators\cite{LRT_Tstar}---, activated carrier densities, and (if needed) impurity in-gap states. For the examples FeSb$_2$, RuSb$_2$, FeAs$_2$ and FeGa$_3$, all characteristic features across transport observables were well captured, including trends between the different compounds.
We therefore believe our phenomenology to be prototypical in the wider context of narrow-gap semiconductors\cite{NGCS}.

\begin{acknowledgments}
The authors gratefully acknowledge discussions with 
R.\ Arita,
N.\ Berlakovich,
G.\ Kotliar, and
Wenhu Xu.
This work has been supported by the Austrian Science Fund (FWF)
through project {\texttt{\textsc{LinReTraCe}}} 
P~30213-N36. 
Calculations were 
performed on the Vienna Scientific Cluster (VSC).
\end{acknowledgments}

\appendix

\section{Transport Kernels}
\label{sec:app_kernels}
In this section we illustrate the contour integration and Matsubara summations necessary to evaluate the transport kernels explicitly. As in the main text we will restrict ourselves to the intra-band case of Eqs.~(\ref{K1}-\ref{K2}). For the sake of brevity we will abridge the notation throughout the derivation and drop the full momentum and band dependence of the spectral function $A(\omega)$, quasi-particle weight $Z$, scattering rate $\Gamma$ and energy $a=\epsilon-\mu$.

\begin{figure*}[th]
  \centering
  \subfloat[$\II_{11}\left(\mathbf{k},n;\Omega;i\nu_m>0\right)$]{
    \includegraphics[width=0.47\textwidth]{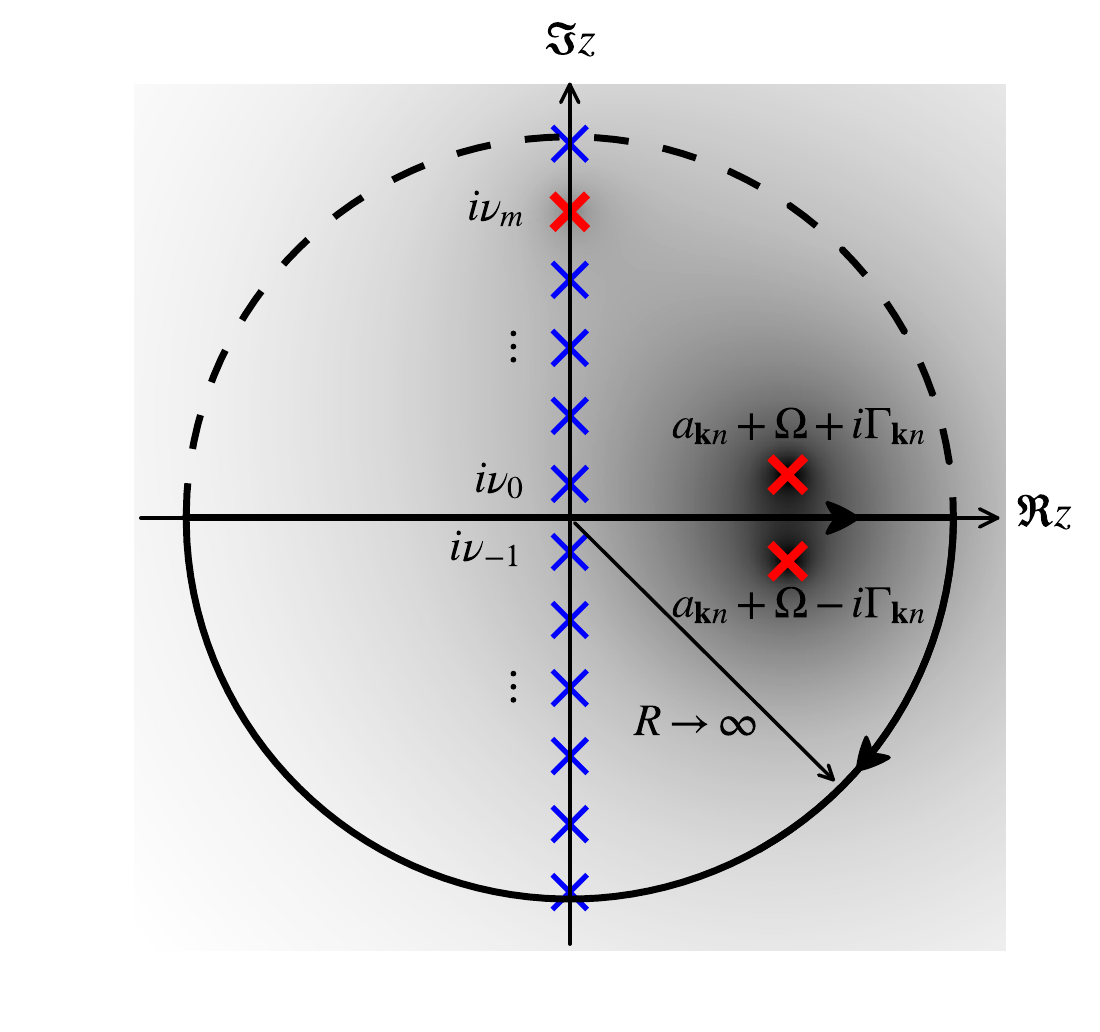}
    }
  \quad
  \subfloat[$\II_{11}\left(\mathbf{k},n;\Omega;i\nu_m<0\right)$]{  \includegraphics[width=0.47\textwidth]{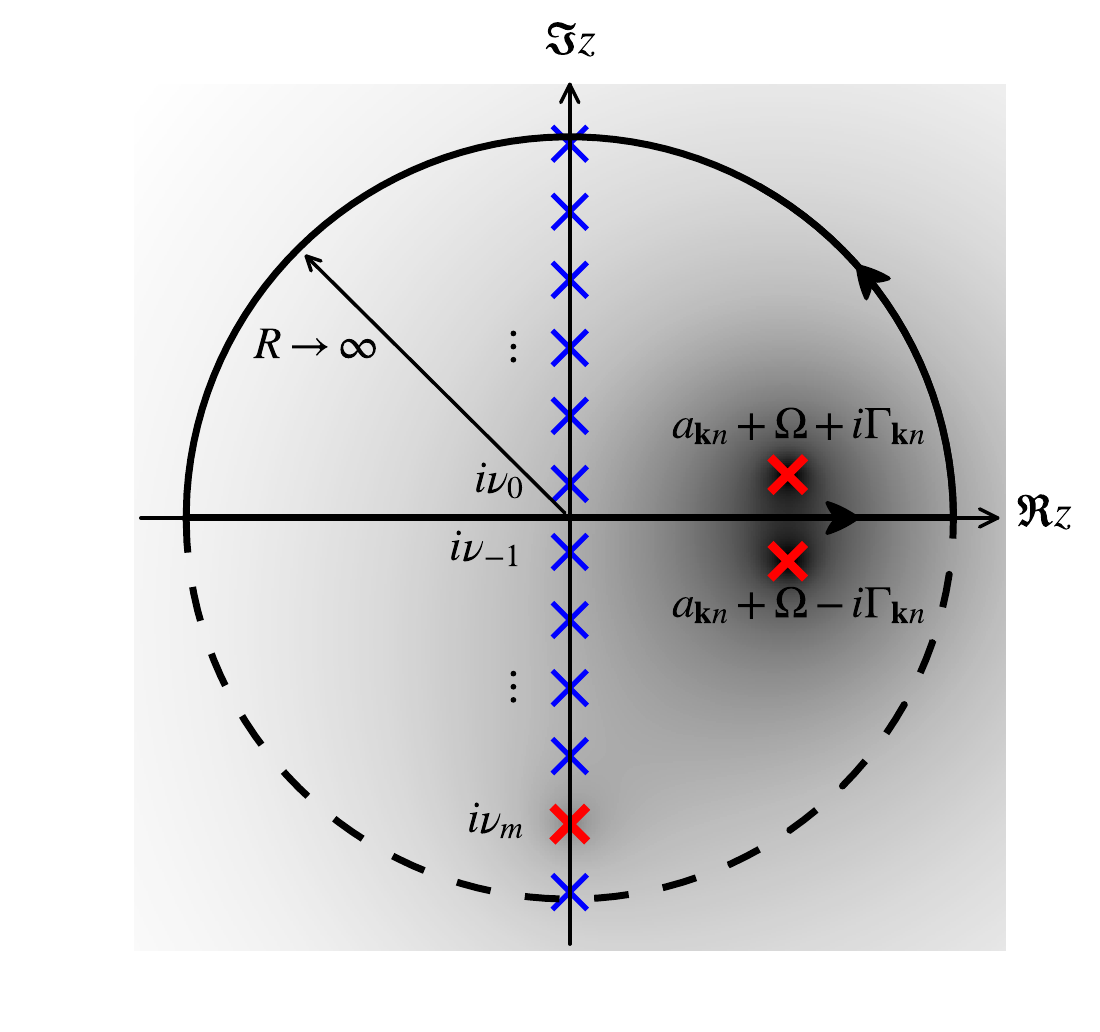}
 }
    \caption{\textbf{Contour integration.} a) Extending the desired integral along $\omega=\Re z$ ($\Im z = 0$), the contour is closed via the \emph{lower} half-plane for $\nu_m >0$ or b) closed via the \emph{upper} half-plane for $\nu_m <0$.
    The poles of the Fermi function are located on the imaginary axis $z=i\nu_m =i\frac{\pi}{\beta}(2m+1)$ while the poles of the spectral function are located at $z=a_{\mathbf{k}n}+\Omega \pm i\Gamma_{\mathbf{k}n}$.
    Due to functional decay $f(z) = \mathcal{O}(z^{a+b-7})$ (see text) in the limit of $R\rightarrow\infty$ the half-circles do not contribute. A straight-forward residue evaluation (inside the closed contour) is thus sufficient to calculate the initial integral.}
    \label{fig:contour}
\end{figure*}

\subsection{Contour integration}
Starting from the generalized transport kernels ($a\in\{1,2\}$, $b\in\{1,2\}$) from \eref{K1}
\begin{equation}
    \mathcal{K}_{ab}(\mathbf{k},n) = \int_{-\infty}^{\infty}d\omega\; \omega^{(a+b-2)} \left( - \frac{\partial f}{\partial\omega} \right) A^2(\omega)
\end{equation}
we insert the Matsubara representation of the derivative of Fermi function
\begin{equation}
\begin{aligned}
    -\frac{\partial f}{\partial\omega} &= \lim_{\Omega\to 0^+} \frac{f(\omega)-f(\omega+\Omega)}{\Omega} \\
    &=  \lim_{\Omega\to 0^+}  \frac{1}{\beta} \sum_m \frac{1}{\Omega} \left(\frac{1}{i\nu_m-\omega} - \frac{1}{i\nu_m-\omega-\Omega}\right)
\end{aligned}
\end{equation}
as well as the (coherent part of the) spectral function
\begin{equation}
    A_{\mathbf{k}n}(\omega) = \frac{Z\Gamma}{\pi} \frac{1}{\left(\omega-a\right)^2 + \Gamma^2}
\end{equation}
where the fermionic Matsubara frequencies are
    $\nu_m = (2m+1)\frac{\pi}{\beta}$ with $m\in\mathbb{Z}$.
The resulting expression
\begin{equation}
\begin{aligned}
    \mathcal{K}_{ab}(\mathbf{k},n) &= \int_{-\infty}^{\infty}d\omega\;  \frac{Z^2\Gamma^2}{\pi^2} \frac{\omega^{(a+b-2)}}{\left[\left(\omega-a\right)^2 + \Gamma^2\right]^2} \;\;\times \\
    & \frac{1}{\beta} \sum_m \lim_{\Omega\to 0^+} \frac{1}{\Omega} \left(\frac{1}{i\nu_m-\omega} - \frac{1}{i\nu_m-\omega-\Omega}\right)    
\end{aligned}
\end{equation}
can be abbreviated with
\begin{equation}
    \mathcal{I}_{ab}(\mathbf{k},n;\Omega;i\nu_m) = \int_{-\infty}^{\infty}d\omega\;  \frac{\left(\omega-\Omega\right)^{(a+b-2)}}{\left[\left(\omega-a-\Omega\right)^2 + \Gamma^2\right]^2} \frac{1}{i\nu_m-\omega}
\label{eq:intab}
\end{equation}
as
\begin{equation}
\begin{aligned}
    \mathcal{K}_{ab}(\mathbf{k},n) &= \frac{Z^2\Gamma^2}{\pi^2} \frac{1}{\beta} \times  \\
   \lim_{\Omega\to 0^+} & \Bigg[\frac{1}{\Omega}
    \sum_m \left( \II_{ab}(\mathbf{k},n;0;i\nu_m) - \II_{ab}(\mathbf{k},n;\Omega;i\nu_m) \right) \Bigg]
\end{aligned}
\end{equation}

For finite (positive) scattering rates $\Gamma > 0$ Eq.~\eqref{eq:intab} is an integral over a function with three distinct poles in the complex plane:
$z=a + \Omega + i\Gamma$, $z=a +\Omega - i\Gamma$, and $z=i\nu_m$.
This function decays with $z\to\infty$ as $\mathcal{O}(z^{a+b-7})$
which for all considered values for $a$ and $b$ is stronger than $\mathcal{O}(z^{-2})$ ensuring that any infinitely large arc in the complex plain will have no contribution. Our desired integral, located on the real axis, can therefore be extended to a closed loop and thus be expressed as a sum of residues, see Fig.~\ref{fig:contour}. By always choosing the half-circle opposite to the pole of the Matsubara frequency we can restrict the evaluation to exactly one (higher order) pole
\begin{equation}
\begin{aligned}
    \mathcal{I}_{ab}(\mathbf{k},n;\Omega;i\nu_m)
    =& \oint_\mathcal{C} dz\;  \frac{\left(z-\Omega\right)^{(a+b-2)}}{\left[\left(z-a-\Omega\right)^2+\Gamma^2\right]^2} \frac{1}{i\nu_m-z} \\
    =& -\sign (\nu_m)\;2\pi i \;\mathrm{Res}_{z=(a+\Omega-i\sign (\nu_m)\Gamma)} \\ &\frac{\left(z-\Omega\right)^{(a+b-2)}}{\left[\left(z-a-\Omega\right)^2+\Gamma^2\right]^2} \frac{1}{i\nu_m-z}.
\end{aligned}
\end{equation}
Due to the different mathematical integration directions positive and negative fermionic Matsubara frequencies result in differing signs.

Evaluating the residue at  $z=(a+\Omega-i\sign (\nu_m)\Gamma)$ results in the following  expressions

\begin{widetext}
\begin{align}
&\mathcal{I}_{11}(\mathbf{k},n;\Omega;i\nu_m) = \frac{\pi}{2\Gamma^3} \left[\frac{i\Gamma\sign(\nu_m)}{\left[i\nu_m -a-\Omega + i\Gamma\sign(\nu_m)\right]^2} + \frac{1}{\left[i\nu_m -a -\Omega+ i\Gamma\sign(\nu_m)\right]} \right]\label{eq:i11} \\
&\begin{aligned}
\mathcal{I}_{12}(\mathbf{k},n;\Omega;i\nu_m) =& \frac{\pi}{2\Gamma^3} \left[ \frac{\Gamma^2 + i\left(a+\Omega\right)\Gamma\sign(\nu_m)}{\left[i\nu_m -a-\Omega + i\Gamma\sign(\nu_m)\right]^2} + \frac{\left(a+\Omega\right)}{\left[i\nu_m -a-\Omega + i\Gamma\sign(\nu_m)\right]}\right] 
% \\
- \Omega\mathcal{I}_{11}(\mathbf{k},n;\Omega;i\nu_m)\label{eq:i12}
\end{aligned}\\
&\begin{aligned}
\mathcal{I}_{22}(\mathbf{k},n;\Omega;i\nu_m) &= \frac{\pi}{2\Gamma^3} \left[ \frac{i\left(a+\Omega\right)^2\Gamma\sign(\nu_m) + 2\left(a+\Omega\right)\Gamma^2 - i\Gamma^3\sign(\nu_m)}{\left[i\nu_m-a-\Omega+i\Gamma\sign(\nu_m)\right]^2} + \frac{\Gamma^2 + \left(a+\Omega\right)^2}{\left[i\nu_m-a-\Omega+i\Gamma\sign(\nu_m)\right]} \right] \\
&-2\Omega\mathcal{I}_{12}(\mathbf{k},n;\Omega;i\nu_m) +\Omega^2 \mathcal{I}_{11}(\mathbf{k},n;\Omega;i\nu_m).\label{eq:i22}
\end{aligned}
\end{align}
\end{widetext}

\subsection{Matsubara sums}
The second step is to perform the Matsubara sums
\begin{equation}
    \II_{ab}(\mathbf{k},n;\Omega) = \frac{1}{\beta} \sum_{m=-\infty}^\infty \II_{ab}(\mathbf{k},n;\Omega;i\nu_m).
\end{equation}
Using the series representation of the digamma and polygamma functions
\begin{align}
    \psi(z) &= -\gamma + \sum_{n=1}^\infty \left(\frac{1}{n} - \frac{1}{n+z}\right) \\
    \psi_{m>0}(z) &= \left(-1\right)^{m+1} m! \sum_{k=0}^\infty \frac{1}{\left( z+k \right)^{m+1}},
\end{align}
the summations appearing in Eqs.~(\ref{eq:i11}-\ref{eq:i22}) result in
\begin{align}
    \frac{1}{\beta} \sum_{m=-\infty}^\infty \frac{1}{i\nu_m - a + i\Gamma\sign(\nu_m)} &= -\frac{1}{\pi} \Im\psi(z) \\
    \frac{1}{\beta} \sum_{m=-\infty}^\infty \frac{1}{\left[i\nu_m - a + i\Gamma\sign(\nu_m)\right]^2} &= -\frac{\beta}{2\pi^2} \Re\psi_1(z) \\
    \frac{1}{\beta} \sum_{m=-\infty}^\infty \frac{\sign(\nu_m)}{\left[i\nu_m - a + i\Gamma\sign(\nu_m)\right]^2} &= -\frac{i\beta}{2\pi^2} \Im\psi_1(z) \\
    \frac{1}{\beta} \sum_{m=-\infty}^\infty \frac{1}{\left[i\nu_m - a + i\Gamma\sign(\nu_m)\right]^3} &= \frac{\beta^2}{8\pi^2} \Im\psi_2(z) \\
    \frac{1}{\beta} \sum_{m=-\infty}^\infty \frac{\sign(\nu_m)}{\left[i\nu_m - a + i\Gamma\sign(\nu_m)\right]^3} &= -\frac{i\beta^2}{8\pi^3} \Re\psi_2(z)
\end{align}
with $z=\frac{1}{2} + \frac{\beta}{2\pi}\left(\Gamma + ia\right)$. Then, the transport integrals simplify to
\begin{widetext}
\begin{align}
    &\mathcal{I}_{11}(\mathbf{k},n;\Omega) = \frac{\pi}{2\Gamma^3} \left[\frac{\beta\Gamma}{2\pi^2}\Im\psi_1 \left(z+\frac{i\beta\Omega}{2\pi}\right) -\frac{1}{\pi}\Im\psi\left(z+\frac{i\beta\Omega}{2\pi}\right)\right] \\
&\begin{aligned}
    \mathcal{I}_{12}(\mathbf{k},n;\Omega) =& \frac{\pi}{2\Gamma^3} \left[ -\frac{\Gamma^2\beta}{2\pi^2}\Re\psi_1\left(z+\frac{i\beta\Omega}{2\pi}\right)
    +\frac{\left(a+\Omega\right)\Gamma\beta}{2\pi^2}\Im\psi_1\left(z+\frac{i\beta\Omega}{2\pi}\right)
    -\frac{\left(a+\Omega\right)}{\pi}\Im\psi\left(z+\frac{i\beta\Omega}{2\pi}\right) \right] \\
    -& \Omega\mathcal{I}_{11}(\mathbf{k},n;\Omega)     
\end{aligned}\\
&\begin{aligned}
    \mathcal{I}_{22}(\mathbf{k},n;\Omega) &= \frac{\pi}{2\Gamma^3} \Bigg[
    \frac{\left(a+\Omega\right)^2\Gamma\beta-\Gamma^3\beta}{2\pi^2}\Im\psi_1\left(z+\frac{i\beta\Omega}{2\pi}\right)
    -\frac{2\left(a+\Omega\right)\Gamma^2\beta}{2\pi^2}\Re\psi_1\left(z+\frac{i\beta\Omega}{2\pi}\right) \\
    &-\frac{\Gamma^2+\left(a+\Omega\right)^2}{\pi}\Im\psi\left(z+\frac{i\beta\Omega}{2\pi}\right) \Bigg] \\
    &-2\Omega\mathcal{I}_{12}(\mathbf{k},n;\Omega) +\Omega^2 \mathcal{I}_{11}(\mathbf{k},n;\Omega).    
\end{aligned}
\end{align}
\end{widetext}

\subsection{Dynamic limit}
Taylor expanding the frequency-dependent di- and polygamma ($m>0$) functions around $z$
\begin{equation}
    \psi\left(z+\frac{i\beta\Omega}{2\pi}\right) = \psi\left(z\right) + \frac{i\beta\Omega}{2\pi}  \psi_{1}\left(z\right) + \mathcal{O}\left(\Omega^2\right)     
\end{equation}
\begin{equation}
    \psi_m\left(z+\frac{i\beta\Omega}{2\pi}\right) = \psi_m\left(z\right) + \frac{i\beta\Omega}{2\pi}  \psi_{m+1}\left(z\right) + \mathcal{O}\left(\Omega^2\right)     
\end{equation}
and evaluating the limit
\begin{equation}
    \KK_{ab}(\mathbf{k},n) = \frac{Z^2\Gamma^2}{\pi^2} \lim_{\Omega\to 0^+} \frac{1}{\Omega} \left[\II_{ab}(\mathbf{k},n;0) - \II_{ab}(\mathbf{k},n;\Omega) \right]
\end{equation}

we finally arrive at the intra-band equations listed in the main text

\begin{widetext}
\begin{align}
\mathcal{K}_{11}(\mathbf{k},n) &= \frac{Z^2\beta}{4\pi^3\Gamma} \left[ \Re\psi_1(z) - \frac{\beta\Gamma}{2\pi} \Re\psi_2(z) \right] \\
\mathcal{K}_{12}(\mathbf{k},n) &= \frac{Z^2\beta}{4\pi^3\Gamma} \left[ a\Re\psi_1(z) - \frac{a\Gamma\beta}{2\pi} \Re\psi_2(z) -\frac{\Gamma^2\beta}{2\pi} \Im\psi_2(z) \right]\\
\mathcal{K}_{22}(\mathbf{k},n) &= \frac{Z^2\beta}{4\pi^3\Gamma} \Bigg[(a^2+\Gamma^2) \Re\psi_1(z) +\frac{\beta}{2\pi}\Gamma\left(\Gamma^2-a^2\right) \Re\psi_2(z)  
- \frac{\beta}{\pi} a\Gamma^2\Im\psi_2(z)
 \Bigg].
\end{align}
\end{widetext}

The equivalent calculation is easily performed for the magnetic kernel functions
\begin{equation}
    \mathcal{K}^B_{ab}(\mathbf{k},n) =  \int_{-\infty}^{\infty}d\omega\; \omega^{(a+b-2)} \left( - \frac{\partial f}{\partial\omega} \right) A^3_{\mathbf{k}n}(\omega)
\end{equation}
which only differ from their non-magnetic counterparts by an additional spectral function. The integrals
\begin{equation}
    \mathcal{I}^B_{ab}(\mathbf{k},n;\Omega;i\nu_m) = \int_{-\infty}^{\infty}d\omega\;  \frac{\left(\omega-\Omega\right)^{(a+b-2)}}{\left[\left(\omega-a-\Omega\right)^2 + \Gamma^2\right]^3} \frac{1}{i\nu_m-\omega}
\end{equation}
then evaluate to
\begin{widetext}
\begin{align}
&\begin{aligned}
\mathcal{I}^B_{11}(\mathbf{k},n;\Omega;i\nu_m) = \frac{i\pi}{8\Gamma^5} &\Bigg[ \frac{2i\Gamma^2}{\left[i\nu_m-a-\Omega+i\Gamma\sign(\nu_m)\right]^3} +\frac{3\Gamma\sign(\nu_m)}{\left[i\nu_m-a-\Omega+i\Gamma\sign(\nu_m)\right]^2} \\
& + \frac{-3i}{\left[i\nu_m-a-\Omega+i\Gamma\sign(\nu_m)\right]} \Bigg]
\end{aligned}\\
&\begin{aligned}
\mathcal{I}^B_{12}(\mathbf{k},n;\Omega;i\nu_m) = \frac{i\pi}{8\Gamma^5}  &\Bigg[ \frac{2i\left(a+\Omega\right)\Gamma^2 + 2\Gamma^3\sign(\nu_m)}{\left[i\nu_m-a-\Omega+i\Gamma\sign(\nu_m)\right]^3} 
+\frac{3\left(a+\Omega\right)\Gamma\sign(\nu_m) - i\Gamma^2}{\left[i\nu_m-a-\Omega+i\Gamma\sign(\nu_m)\right]^2}
\\ &+ \frac{-3i\left(a+\Omega\right)}{\left[i\nu_m-a-\Omega+i\Gamma\sign(\nu_m)\right]} \Bigg]
- \Omega \mathcal{I}^B_{11}(\mathbf{k},n;\Omega;i\nu_m)
\end{aligned}\\
&\begin{aligned}
\mathcal{I}^B_{22}(\mathbf{k},n;\Omega;i\nu_m) = \frac{i\pi}{8\Gamma^5} &\Bigg[ \frac{2i\left(a+\Omega\right)^2\Gamma^2 + 4\left(a+\Omega\right)\Gamma^3\sign(\nu_m) - 2i\Gamma^4}{\left[i\nu_m-a-\Omega+i\Gamma\sign(\nu_m)\right]^3} \\
&+\frac{-2i\left(a+\Omega\right)\Gamma^2 + \Gamma^3\sign(\nu_m)+3\left(a+\Omega\right)^2\Gamma\sign(\nu_m)}{\left[i\nu_m-a-\Omega+i\Gamma\sign(\nu_m)\right]^2} \\
&+ \frac{-i\Gamma^2 - 3i\left(a+\Omega\right)^2}{\left[i\nu_m-a-\Omega+i\Gamma\sign(\nu_m)\right]} \Bigg] - 2\Omega \mathcal{I}^B_{12}(\mathbf{k},n;\Omega;i\nu_m) + \Omega^2 \mathcal{I}^B_{11}(\mathbf{k},n;\Omega;i\nu_m).
\end{aligned}
\end{align}
\end{widetext}
Performing the Matsubara sums and taking the dynamical limit, the transport kernels results in the magnetic transport kernels
\begin{widetext}
\begin{align}
&\mathcal{K}^B_{11}(\mathbf{k},n) = \frac{Z^3\beta}{16\pi^4\Gamma^2} \left[ 3 \Re\psi_1(z) - \frac{3\Gamma\beta}{2\pi} \Re\psi_2(z) + \frac{\Gamma^2\beta^2}{4\pi^2}\Re\psi_3(z) \right]\\
&\mathcal{K}^B_{12}(\mathbf{k},n) = \frac{Z^3\beta}{16\pi^4\Gamma^2} \Bigg[ 3a \Re\psi_1(z) - \frac{3a\Gamma\beta}{2\pi} \Re\psi_2(z) - \frac{\Gamma^2\beta}{2\pi} \Im\psi_2(z) + \frac{a\Gamma^2\beta^2}{4\pi^2}\Re\psi_3(z) + \frac{\Gamma^3\beta^2}{4\pi^2} \Im\psi_3(z)\Bigg] \\
&\begin{aligned}
\mathcal{K}^B_{22}(\mathbf{k},n) = \frac{Z^3\beta}{16\pi^4\Gamma^2} \Bigg[& (\Gamma^2+3a^2) \Re\psi_1(z)  - \frac{\beta\Gamma(\Gamma^2+3a^2)}{2\pi} \Re\psi_2(z) - \frac{\beta a\Gamma^2}{\pi} \Im\psi_2(z) \\
&-\frac{\beta^2\Gamma^2(\Gamma^2-a^2)}{4\pi^2}\Re\psi_3(z) + \frac{\beta^2 a \Gamma^3}{2\pi^2} \Im\psi_3(z) \Bigg].
\end{aligned}
\end{align}
\end{widetext}

\section{Low temperature expansion}
\label{sec:app_lowT}

In order to evaluate the low temperature behavior we exploit the Taylor expansion around $\overline{z}=\infty$ of the digamma function
\begin{equation}
    \Psi\left(\frac{1}{2}+\overline{z}\right) = \ln(\overline{z}) + \frac{1}{24 \overline{z}^2} - \frac{7}{960 \overline{z}^4} + \mathcal{O}\left(\overline{z}^{-6}\right),
\end{equation}
and all higher order polygamma functions
\begin{align}
    \Psi_1\left(\frac{1}{2}+\overline{z}\right)&= \frac{1}{\overline{z}}-\frac{1}{12 \overline{z}^3}+\frac{7}{240 \overline{z}^5}+\mathcal{O}\left(\overline{z}^{-7}\right) \\
    \Psi_2\left(\frac{1}{2}+\overline{z}\right)&=-\frac{1}{\overline{z}^2}+\frac{1}{4 \overline{z}^4}-\frac{7}{48 \overline{z}^6} +\mathcal{O}\left(\overline{z}^{-8}\right) \\
    \Psi_3\left(\frac{1}{2}+\overline{z}\right)&=\frac{2}{\overline{z}^3}-\frac{1}{\overline{z}^5}+\frac{7}{8 \overline{z}^7} + \mathcal{O}\left(\overline{z}^{-9}\right).
\end{align}
Eqs.~(\ref{LRT_k11}-\ref{LRT_k12B}) then become
\begin{widetext}
\begin{eqnarray}
\mathcal{K}_{11}&=&\frac{Z^2}{\pi^2}\frac{\Gamma^2}{(a^2+\Gamma^2)^2} \times \left[ 1 + \frac{2\pi^2}{3}\frac{5a^2-\Gamma^2}{(a^2+\Gamma^2)^2}\times k_B^2 T^2 + \mathcal{O}(T^4) \right] \\ 
\mathcal{K}_{12}&=&\frac{4Z^2}{3}\frac{a\Gamma^2}{(a^2+\Gamma^2)^3}\times  \left[ k_B^2T^2 + \frac{7\pi^2}{5}\frac{5a^2-3\Gamma^2}{(a^2+\Gamma^2)^2} \times k_B^4T^4  +\mathcal{O}(T^6)  \right] \\
\mathcal{K}_{22}&=&\frac{Z^2}{3}\frac{\Gamma^2}{(a^2+\Gamma^2)^2} \times \left[ k_B^2T^2 + \frac{14\pi^2}{5}\frac{5a^2-\Gamma^2}{(a^2+\Gamma^2)^2} \times k_B^4T^4   +\mathcal{O}(T^6)   \right]\\
\mathcal{K}^B_{11}&=&\frac{Z^3}{\pi^3}\frac{\Gamma^3}{(a^2+\Gamma^2)^3}
\times \left[ 1 + \pi^2\frac{7a^2-\Gamma^2}{(a^2+\Gamma^2)^2}
\times k_B^2 T^2  +\mathcal{O}(T^4)\right] \\
\mathcal{K}^B_{12}&=& \frac{2Z^3}{\pi} \frac{  a \Gamma^3}{(a^2+\Gamma^2)^4} \times \left[ k_B^2T^2 +
\frac{28 \pi ^2}{15}\frac{ 7 a^2-3 \Gamma^2}{ \left(a^2+\Gamma^2\right)^2} \times k_B^4T^4
+\mathcal{O}(T^6)  \right]\\
\mathcal{K}^B_{22}&=& \frac{Z^3}{3\pi}\frac{\Gamma^3}{\left(a^2+\Gamma^3\right)^3} \times \left[ k_B^2 T^2 + \frac{21\pi^2}{5} \frac{7a^2-\Gamma^2}{\left(a^2+\Gamma^2\right)^2} \times k_B^4 T^4 + \mathcal{O}(T^6)\right]
\label{eq:LlowT}
\end{eqnarray}
\end{widetext}

Therefore for any $\lim_{T\to 0} \Gamma(T) > 0$ the resistivity as well as the Hall coefficient will saturate. Furthermore the Seebeck coefficient, the Nernst coefficient as well as the thermal conductivity will tend to $0$ in a linear fashion.

\section{Kernel approximations}
\label{sec:app_approx}
\begin{figure}[!ht]
    \centering
    \subfloat{{\includegraphics[width=0.45\textwidth]{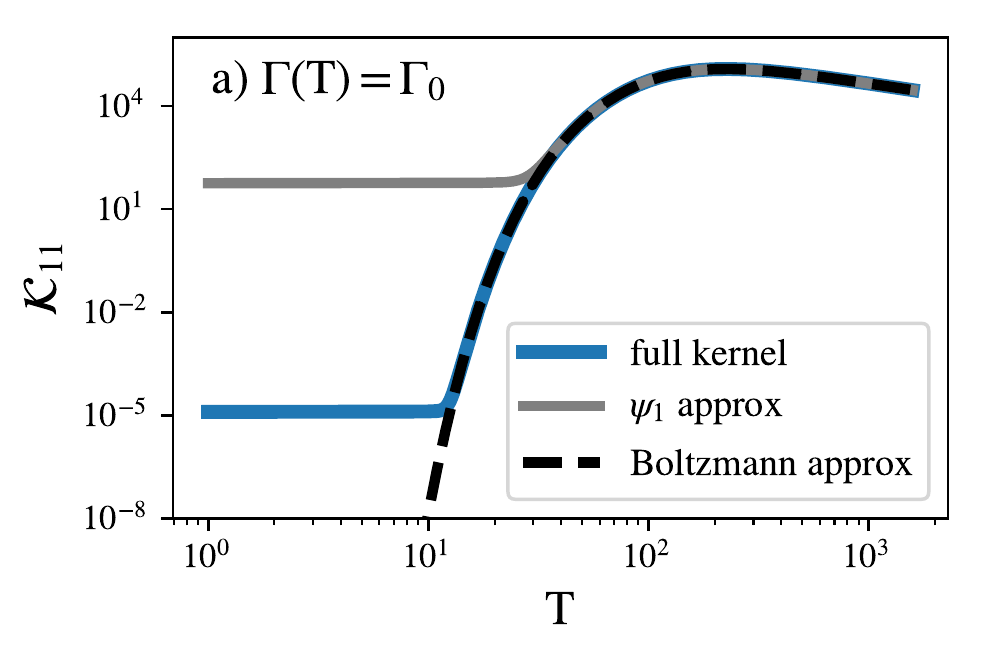}}}
    
    \subfloat{{\includegraphics[width=0.45\textwidth]{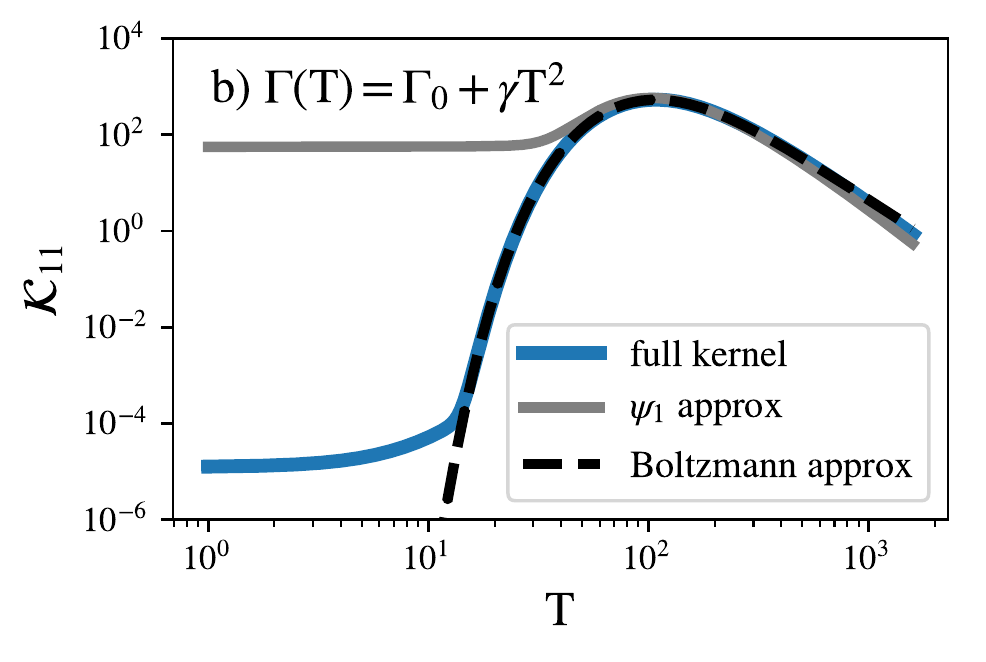}}}
    \caption{$\psi_1$-approximation and Boltzmann approximation of the full kernel expression $\KK_{11}$ for $Z=1$, $a=0.03$, $\Gamma_0=10^{-5}$ and $\gamma=10^{-7}$. a) Kernel behavior with temperature independent scattering rate $\Gamma(T)=\Gamma_0$. b) Kernel behavior with Fermi-liquid like scattering rate $\Gamma(T)=\Gamma_0+\gamma T^2$. Using the lowest order polygamma function is a good approximation at high temperature even for relatively large scattering rates. At low enough temperatures, we find a crossover where both terms of the kernel contribute equally. Surprisingly, the Boltzmann approximation follows the full kernel expression longer, however does not saturate.}
    \label{fig:kernel_approx}
\end{figure}

In order to better understand the transport kernels, we consider two types of approximations to the full kernel expressions: The $\psi_1$-approximation---as the name suggests---consists of simply using the lowest \emph{explicit} order in $\Gamma$, resulting in expressions that functionally depend only on the first order polygamma function, e.g.,
\begin{equation}
    \mathcal{K}^{\psi_1}_{11}(\mathbf{k},n) = \frac{Z^2\beta}{4\pi^3\Gamma} \left[ \Re\psi_1(z) \underbrace{- \frac{\beta\Gamma}{2\pi} \Re\psi_2(z)}_{\equiv\;0} \right].
\end{equation}
Due to the implicit $\Gamma$-dependence through the polygamma argument, $z=\frac{1}{2}+\frac{\beta}{2\pi}(\Gamma+i a)$, this approximation is not reasonable in the zero temperature limit where the $\psi_2$-term become equally important, see \aref{sec:app_lowT}.
The second approximation goes one step further and also applies the limit of $\Gamma\to 0^+$ to the remaining $\psi_1$-function, recovering the Boltzmann expression
\begin{equation}
    \mathcal{K}^{\mathrm{Boltzmann}}_{11}(\mathbf{k},n) = \frac{Z^2\beta}{4\pi^3\Gamma} \left[ \underbrace{\Re\psi_1(z)}_{\equiv -\frac{2\pi^2}{\beta}f^{\prime}(a)} \underbrace{- \frac{\beta\Gamma}{2\pi} \Re\psi_2(z)}_{\equiv\;0} \right].
\end{equation}
The full kernel and the corresponding approximations, illustrated in \fref{fig:kernel_approx}, show that the $\psi_1$-approximation produces a finite saturation value in the zero temperature limit--that, however, emerges at too large $T$. Surprisingly, the more restrictive $\Gamma\to 0$ Boltzmann approximation (dashed) leads to \emph{better} agreement at intermediate temperatures. Yet, this comes with a trade-off: the Boltzmann kernel is able to trace the full kernel down to lower temperature, but does not saturate: $\KK_{11}^\mathrm{Boltzmann} (T\rightarrow 0) = 0$. Identical behaviors can be observed for the higher order transport kernels.
Combined to observable transport tensors this deficiency in the limiting behavior causes the usual problem of the relaxation time approximation with its nonphysical entropy transport in gapped systems. Employing the full kernels on the other hand does not suffer from this problem and entropy-transport quantities become  thermodynamically consistent.

\section{Scaling with lattice constants and band structure}
\label{sec:app_latticescaling}
Due to the reductionist approach in the main text (only nearest-neighbor hopping, cubic lattice constant $a_{\mathrm{lattice}}=1$\AA) the absolute values of the transport properties were for the most part ignored. A first step towards more realism is using a proper lattice spacing $a_{\mathrm{lattice}}$: The (cubic) unit-cell volume is $V = a_{\mathrm{lattice}}^3$, while the optical elements within the Peierls approximation (Sec.~\ref{Peierls}) scale with $M \propto a_{\mathrm{lattice}}^2$ and $M^B \propto a_{\mathrm{lattice}}^4$. Combined, the Onsager coefficients scale like 
\begin{align}
    \LL_{ab} &\propto a_{\mathrm{lattice}}^{-1} \\
    \LL^{B}_{ab} &\propto a_{\mathrm{lattice}}.
\end{align}
Eqs.~(\ref{eq:ObservCond}-\ref{eq:ObservNernst}) then naturally lead to
\begin{align}
    \rho &\propto a_{\mathrm{lattice}} \\
    S &\equiv \mathrm{const} \\
    R_H &\propto a_{\mathrm{lattice}}^3 \\
    \nu &\propto a_{\mathrm{lattice}}^2.
\end{align}

The second step is a more realistic band structure and transition matrix elements.
An input from a density functional calculations and including beyond-Peierls matrix elements is straight-forward within \verb=LinReTraCe=\cite{LRT}. Here, however, we wanted to restrict ourselves to a tight-binding model with hoppings $t$ to extract the most essential physics. Then, matrix elements transform like $M \propto t^2$ and $M^B \propto t^3$.
In a multi-band situation, an overall scaling of results for varying hopping parameters is not possible.

\end{document}